\documentstyle[12pt,epsfig]{article}
\newcommand{\beq}{\begin{equation}}
\newcommand{\eeq}{\end{equation}}
\newcommand{\bea}{\begin{eqnarray}}
\newcommand{\eea}{\end{eqnarray}}

\def\N{{\scriptscriptstyle N}}

\font\titre=cmssdc10 scaled 1440

\font\rmtit=cmr12

\newcommand{\bce}{\begin{center}}
\newcommand{\ece}{\end{center}}

\def\kv{{\vec k}}
\def\Kv{{\vec K}}
\def\pv{{\vec p}}
\def\qv{{\vec q}}

\def\Rv{{\vec R}}
\def\cale{{\cal E}}
\def\PR{{\cal P}(R)}
\def\epsF{{\varepsilon_F}}

\begin{document}

\title{\hfill {\rmtit DFTT 39/1997}\\
\vspace*{1.cm}{\titre
{\bf The semi--classical approach to the exclusive electron
scattering}}}
\author{ W.M. Alberico$^{(1)}$, G. Chanfray$^{(2)}$, 
J. Delorme$^{(2)}$, \\
M. Ericson$^{(2/3)}$, A. Molinari$^{(1)}$
\vspace*{0.3cm}\\
{\it $^{(1)}$ Dipartimento di Fisica Teorica, Universit\`a di
Torino}\\
{\it  and INFN, Sezione di Torino, via P. Giuria 1, Torino, Italy} \\
{\it $^{(2)}$ Institut de Physique Nucl\'eaire de Lyon, IN2P3--CNRS}\\
{\it et Universit\'e Claude Bernard, F 69622 Villeurbanne Cedex, 
France}\\
{\it $^{(3)}$ CERN, Geneva, Switzerland}}
\maketitle


\vskip 1.5 cm

\begin{abstract}
The semiclassical approach, successfully applied in the past to the 
inelastic, inclusive electron scattering off nuclei, is extended to the 
treatment of exclusive processes.  
The final states interaction is accounted for in the mean field approximation,
respecting the Pauli principle. The impact on the exclusive cross section
of the shape of the potential binding the nucleons into the nucleus and of the
distortion of the outgoing nucleon wave are explored. 
The exclusive scattering is found to be quite sensitive to the mean field
final states interaction, unlike the inclusive one. Indeed we verify that
the latter is not affected, as implied by unitarity, by the distortion of
the outgoing nucleon wave except for the effect of relativity, which is
modest in the range of momenta up to about $500$~MeV/c.
Furthermore, depending upon the correlations between the directions of the 
outgoing and of the initial nucleon, the exclusive cross--section 
turns out to be remarkably 
sensitive to the shape of the potential binding the nucleons. 
These correlations also critically affect the domain in the missing energy--
missing momentum plane where the exclusive process occurs.
\end{abstract}

\section{Introduction}
\setcounter{equation}{0}
The plane wave impulse approximation (PWIA) has been a framework extensively
employed in analyzing the exclusive (or semi--inclusive)
processes of inelastic scattering of electrons off nuclei, like, e.g., the
$(e,e'p)$ one\cite{Day}.
The advantage of such an approach lies of course in its simplicity~: indeed, 
in the PWIA, the final nucleon is described by a plane wave and is not 
antisymmetrized with the daughter ($A-1$)
nucleus. Accordingly in PWIA one deals with
the diagonal component of the spectral function $S(\vec p, E_{ p})$ only,
which is the easiest to calculate.

In this connection the Fermi gas (FG) model,
where translational invariance forces the outgoing nucleon wavefunction to be a
plane wave anyway, appears naturally related, in the non--Pauli blocked
domain, to the PWIA, its spectral function being intrinsically 
diagonal\cite{REF7}:
at the same time, however, the FG spectral function cannot be considered 
as a realistic one, as it stems from a uniform distribution of particles.

The chief flaw of PWIA is of course the neglect of final state 
interactions (FSI). These play an important role also in the FG, where in 
fact they can be more easily treated, although the infinite volume of the
system appears hardly suitable for describing 
exclusive processes. Yet finite size (surface) and binding effects can
be easily inserted into the FG model semi--classically.
Indeed a semi--classical formalism has been developed\cite{REF2,REF3}, 
which starting
from the FG satisfactorily accounts for the impact of the finite size 
and of the binding energy 
of the nucleus in the ``inclusive'' response
functions, but for their low energy side, where the semi--classical 
approach interpolates the quantum mechanical cross sections without 
reproducing their rapid variation with the energy.

The aim of this paper is to extend the semi--classical method to deal 
with exclusive processes, going beyond the PWIA by accounting for the
FSI. Of course the problem of the FSI in the exclusive processes has 
already been widely addressed\cite{REF1}: not, however, in the semi--classical
framework. In this paper we shall show that the latter not only allows 
an adequate treatment of the inclusive processes, but also of 
the exclusive (or semi--inclusive) ones, yielding 
an off--diagonal spectral function $S (\vec p, {\vec p}\,' E)$ 
and a projection operator $\hat \rho_\N$ (to be later defined) 
properly describing (the first) the dynamics of a nucleon inside the 
finite nucleus  and (the second) the distortion the outgoing nucleon 
suffers in crossing the nuclear surface, at least in a mean field 
framework. Indeed this  is the scheme we shall adhere to, 
for simplicity, in the present work: however the extension of the 
semi--classical method to encompass nucleon--nucleon (NN) dynamical 
correlations appears to be entirely 
feasible, both for the spectral function and for the
distortion of the final nucleon wave. We 
actually  intend to carry it out since the data unambiguously point to
the existence of a substantial exclusive cross--section in kinematical 
regions hardly compatible with 
a pure mean field description of the
$(e,e'\,p)$ process\cite{Sick}.

As it is well--known the 
semi--classical method expresses the physical observables in powers of
the Planck's constant $\hbar$. It turns out that the leading 
term of the expansion already accounts, in the mean field
approximation, for the basic elements of 
the exclusive physics, namely the nuclear confinement and the FSI.
This important finding allows then to address several questions related 
to the physics of the exclusive (and inclusive) processes, whose answer
would otherwise be much harder to get in a fully quantum mechanical 
many--body scheme. 

In order to appreciate the impact on the exclusive process of the FSI,
we describe the distortion of the outgoing nucleon wave 
in two opposite, schematic models, the so--called eikonal and 
uniform approximations:
in the first the outgoing nucleon is not deflected 
from the direction of the initial momentum, 
while in the latter the final nucleon is 
isotropically emitted from the nucleus. 
For the same purpose we also felt it useful to evaluate, within 
the semi--classical approach, the exclusive cross--sections 
in the PWIA and to compare it with the results which include FSI.

For the purpose of testing the sensitivity of the semi--classical
exclusive cross--section 
to the shape of the shell model potential binding 
the nucleons in the nucleus, we employ both an harmonic oscillator 
and a Woods--Saxon potential well. We remind that a unified approach, 
where hole and particle states are treated on equal footing, has been 
developed within 
a complex shell model potential which extrapolates the mean field from
positive toward negative energies\cite{Mahaux1,Mahaux2}. Also in the 
present treatment particle and holes states are affected by the same
mean field, although we consider here, for simplicity, only real 
potential wells. 

Indeed in the exclusive $(e, e'\,p)$ cross--section the outgoing 
nucleon wave can keep track of the original bound state, again 
depending upon the distortion mechanism. 
On the contrary, different mean fields are 
known to play a minor role in the inclusive cross--section, which 
turns out to be almost unaffected by the shape of the potential. 

We obtained the latter by integrating the exclusive cross--section
in the appropriate domain of the missing energy -- missing momentum 
plane: due to the approximate treatment of the distortion 
mechanism, it is not ``a priori'' guaranteed that the inclusive 
cross--sections are, as indeed they must in the mean field
approximation exploited here, insensitive to the distortion of 
the outgoing nucleon.

Although the main focus of this work is centered on the effects of FSI,
special attention has been also devoted to the spectral function, which is
a key ingredient of the exclusive cross--section: direct calculation of
the hole spectral density have been performed long ago with variational
methods in few--nucleon systems\cite{Ciofi,Schiavilla} and later on in
nuclear matter\cite{Benhar1,Benhar2} in the frame of the correlated basis
theory. Spectroscopic factors have been recently evaluated in a relativistic
shell model approach\cite{Udias1}, both in medium and heavy nuclei. The 
effect of a relativistic optical potential in exclusive processes is also
investigated, to account for the distortion of the ejected 
nucleon\cite{Udias2}. While the semi--classical approach admittedly 
fails in reproducing specific quantum mechanical effects, we will show that
in the spectral function it retains some important features which are lost
in the framework of a Fermi gas description, still keeping a deal of 
simplicity.

We summarize now the organization of the paper: in Section 2 the
general expressions for the inclusive and exclusive cross--section in the
one--photon exchange approximation are shortly revisited. Furthermore the
off--diagonal spectral function and the distortion operator are introduced.
In Section 3 we derive and discuss the semi--classical expression for the
exclusive cross--section, both in PWIA and DWIA. 
In Section 4 the diagonal part of the 
semi--classical spectral function is obtained in the mean field 
framework. Analytic expressions of this quantity for a few one--body 
potential wells are provided in Appendix A. 
In Section 5 we deduce the mean field expression for 
the distortion operator and  discuss it in the
context of the two rather extreme models (eikonal and uniform approximation)
referred to above. In Section 6  we calculate, for both models, 
the exclusive cross--sections and, in Section 7,
 the inclusive ones, by integrating the former
 over appropriate regions of the missing 
energy--missing momentum plane.
Finally in Section 8 we present and discuss our numerical results, while 
Section 9 illustrates the merits of the semi--classical approach and its 
possible extensions.

\section{The cross--sections}
\setcounter{equation}{0}

The inclusive
cross-section for the scattering of an electron, with initial and final
four-momenta $k$  and $k'$ respectively, out of a nucleus, 
initially in its ground state $\vert A >$ and then excited 
into ``any'' final state $\vert X >$, reads
\beq
\frac{d^2\sigma}{ d\Omega_e d\epsilon'} = 
2 \alpha^2 \frac{1}{ Q^4_{\mu}} \ \frac{k'}{k} \eta^{\mu\nu} \ {\sum_X} 
< A\vert \hat J^\dagger_{\mu} 
(\vec q)\vert \ X ><X  \vert \ \hat J_{\nu} (\vec q)\vert A > 
\delta (E_X - E_A - \omega)\,
\label{II.1}
\eeq
(all the states are normalized to one in a large box of volume V). 
In the above $Q^2_\mu = \omega^2 - {\vec q\,}^2$ is the space--like four 
momentum transferred from the electron to the nucleus and
\beq
\eta^{\mu\nu} = k^{\mu} k'^{\nu} + k^{\nu}k'^{\mu} - 
g^{\mu\nu} k\cdot k'
\label{II.2}
\eeq
is the well--known symmetric leptonic tensor of rank two.
The diagram describing the inclusive process is displayed in Fig. 1.

In the present work we confine ourselves, for sake both of simplicity and of
illustration, to consider one-body current only, disregarding meson exchange 
currents (MEC). Likewise correlations among nucleons beyond the mean field 
will be dealt with in future research. In any case
the matrix elements of the nucleon's one--body current $\hat J_{\nu}$ 
entering into (\ref{II.1})
read (the symbols are self--explanatory)
\beq
< X\vert \hat J_{\nu} (\vec q)\vert A > =\ \sum_{s}
\int \frac{d\vec p }{ (2\pi)^3} \
\int \frac{d{\vec p}\, '} { (2\pi)^3} < X \vert \hat a^\dagger
 ({\vec p}\ ', s')\, \hat a 
(\vec p, s)\vert A >\, <{\vec p}\ ', s'\vert j_{\nu} (\vec q)\vert\vec p,s>\, ,
\label{II.3}
\eeq
the standard annihilation and creation operators $\hat a$ and 
$\hat a^\dagger$ being normalized according to the anticommutation rule
\beq
\lbrace \hat a \ (\vec p, s), \ \hat a^\dagger ({\vec p}\ ', s')\rbrace = 
(2\pi)^3 \delta (\vec p - {\vec p}\ ')\delta_{ss'}\,.
\label{II.7}
\eeq
In the above
\beq
< {\vec p}\, ', s'\vert  j_{\nu} (\vec q)\vert \vec p,s > = (2\pi)^3 \delta 
({\vec p}\,'- \vec p - \vec q) j_{\nu}^{s's} (\vec p + \vec q, \vec p)
\label{II.4}
\eeq
and
\beq
j_{\nu}^{s's} (\vec p + \vec q, \vec p) = \bar u (\vec p + \vec q,s')
\Big\lbrack F_1 (Q^2_{\mu}) \gamma_{\nu} +
 F_2 (Q^2_{\mu}) i \frac{\kappa}{ 2M_{\N}}
\sigma_{\mu\nu} Q^{\mu}\Big\rbrack u (\vec p, s)
\label{II.5}
\eeq
where $F_1$ and $F_2$ are the Dirac and Pauli nucleon's form factors,
$\kappa=1.79$ for protons and $\kappa=-1.91$ for neutrons. 
Finally the spinor normalization is $u^\dagger u = 1$.

Clearly the insertion of (\ref{II.4}) into (\ref{II.3}) leads to
\beq
< X\vert \hat J_{\nu} (\vec q)\vert A > = \sum_{s}
\int \frac{d\vec p}{(2\pi)^3}  
<X\vert \hat a^\dagger (\vec p + \vec q, s') \ \hat a \ (\vec p,s)\vert A > 
j_{\nu}^{s's} (\vec p + \vec q,\vec p).
\label{II.6}
\eeq
Now in the PWIA framework the final nuclear state is factorized as follows
($V$ is the normalization volume)
\beq
\vert X > = 
\vert n > \otimes \frac{1}{\sqrt{V}} \vert \widetilde{\vec p}_\N, s_\N>
\label{II.8}
\eeq
where $\vert n >$ represents an excited state, normalized to one, 
of the residual $(A-1)$ nucleus
and $\vert \widetilde{\vec p}_\N, s_\N >$ is just a 
plane wave, normalized according with the anticommutators (\ref{II.7}). 
In such a scheme, to the matrix element (\ref{II.6})  
only a single nucleon with a given 
momentum $\vec p$ will contribute,
the rest of the nucleus behaving just as a spectator.

In the present approach instead 
the wave of the outgoing nucleon,
$\vert \widetilde{\vec p}_\N, s_\N >$, 
is ``distorted'' by the interaction with the residual nucleus and no 
longer is a momentum eigenstate; as a consequence the nuclear matrix 
element (\ref{II.6}) will be expressed through an integral (sum) over all
possible initial nucleons' momenta (spin) and will read:
\beq
< X \vert \hat J_{\nu} (\vec q) \vert A > \ = \ \frac{1}{\sqrt {V}} \sum_s
\int \frac{d\vec p}{(2\pi)^3} < \widetilde{\vec p}_\N \vert \vec p + 
\vec q > j_{\nu}^{s_\N s} (\vec p + \vec q, \vec p) 
<  n \ \vert \ \hat a \ (\vec p, s) \vert A >
\label{II.9}
\eeq
providing the spin of the outgoing nucleon, which will no 
longer be explicitly 
indicated in the state vectors, is unaffected by the distortion.

In Fig. 2 the exclusive process is diagrammatically displayed~: the bubble 
on the final nucleon leg should be ignored in the PWIA scheme, whereas 
in the framework of the present paper it is meant to embody the FSI.

Using (\ref{II.9}) we can now compute the cross--section for the 
exclusive process. In the laboratory frame, where the nucleus is at rest 
(hence $E_A = M_A$), we obtain
\bea
&&\frac{d^4 \sigma}{d \Omega_e d \epsilon' d \vec p_\N} \ = 
\ \frac{2 \alpha^2}{(2\pi)^3} \ \frac{1}{Q^4_{\mu}} 
\ \frac{k'}{ k} \ \eta_{\mu\nu}
\int \frac{d \vec p}{(2 \pi)^3} \int \frac{d{\vec p}\ '}{(2 \pi)^3}\
\label{II.10}\\
&&\ \times
\sum_{s_\N ,s',s} \ j^{\mu}_{s's_\N} \ ({\vec p}\ ', {\vec p}\ ' + \vec q) 
<{\vec p}\ ' + \vec q\, \vert \,\widetilde{\vec p}_\N >
< \widetilde{\vec p}_\N \,
\vert \, \vec p + \vec q > j^{\nu}_{s_\N\,s} \ (\vec p + \vec q, \vec p)
\nonumber\\
&&\quad\times
\sum_n < A \vert \hat a^\dagger \ ({\vec p}\ ', s')| n > \delta 
\lbrack E^n_{A-1} - (\omega + M_A - E_\N)\rbrack < n \vert \hat a (\vec p,s) 
\vert A >
\nonumber
\eea
the sum extending over the whole spectrum of eigenfunctions of the 
residual (A-1) nucleus, with eigenvalues $E^n_{A-1}$, while
$E_\N$ is the energy of the outgoing nucleon. 

Concerning the electromagnetic vertices they will be taken, according to
(\ref{II.5}), with the nucleon on the mass--shell (which is the case 
for the FG, but clearly not for a finite nucleus). 
Of course this assumption can be corrected
by employing, {\it e.g.}, the CC1  prescription of De Forest \cite{REF4} 
to move the nucleon off the energy shell preserving gauge invariance. 

Let us  now transform the $\delta$--function appearing in (\ref{II.10}). 
For this purpose
it is customary to introduce the positive ``nuclear separation energy''
\beq
E_S = M_\N + M_{A-1} - M_A = M_\N - \mu,
\label{II.11}
\eeq
$\mu$ being the nuclear chemical potential, and the positive ``excitation
energy''
\beq
{\cal E}_n \ = \ E^n_{A-1} - E^0_{A-1} \ \cong \ E^n_{A-1} - M_{A-1} -
E_{\hbox{rec}}
\label{II.12}
\eeq
of the residual (A-1) nucleus recoiling with momentum $\vec q - \vec p_\N$ (the
``missing momentum'') and energy
\beq
E_{\hbox{rec}} \ \cong \ \frac{(\vec q - \vec p_\N)^2}{ 2 \ M_{A-1}}.
\label{II.13}
\eeq
The above formula is valid in the non--relativistic approximation, which is
adequate for the recoil energy.
Accordingly we can write
\beq
\delta \lbrack E^n_{A-1} - (\omega + M_A - E_\N)\rbrack \ = \ \delta \lbrack
{\cal E}_n - (\omega - T_\N - E_S - E_{\hbox{rec}})\rbrack \ = \ 
\delta\left({\cal E}_n - {\cal E}\right)
\label{II.14}
\eeq
where
\beq
{\cal E} \ = \ \omega - T_\N - E_S - E_{\hbox{rec}}
\label{II.15}
\eeq
is the so--called ``missing energy'', fixed by the external kinematics, and 
$T_\N$ the kinetic energy of the outgoing nucleon.

To proceed further we introduce now into the formalism three quantities, 
related to different one--body operators, which allow to express the 
exclusive cross--section in a compact form.

The first of these is the general (off--diagonal) {\em spectral function} 
defined as follows
\bea
< \vec p,  s \vert S ({\cal E}) \vert {\vec p}\ ',s' > 
& =& \sum_n < A \vert  \hat a^\dagger ({\vec p}\ ',s') \vert n > \delta \lbrack
{\cal E} - (E^n_{A-1} - E^0_{A-1})\rbrack < n \vert \hat a (\vec p,s) \vert A>
\nonumber \\
& =& < A \vert \hat a^\dagger ({\vec p}\ ',s) \delta (\widehat{\cal H} - 
{\cal E}) \hat a (\vec p,s) \vert A >.
\label{II.16}
\eea
In the above closure has been applied and $\widehat{\cal H}$
is the Hamiltonian whose eigenvalues are the excitation energies 
of the residual nucleus.

Upon integration over the excitation energy, the spectral function, 
diagonal in the spin indices for parity conserving interactions, yields
\beq
\int d{\cal E}\, < \vec p ,  s \vert S ({\cal E})\vert {\vec p}\ ', s > = 
< A \vert  \,\hat a^\dagger ({\vec p}\ ', s) \hat a (\vec p, s) \vert A >
\label{II.17}
\eeq
which, for $\vec p = {\vec p}\ '$, is just the momentum distribution of 
the nucleons inside the nucleus. Thus the latter becomes 
experimentally accessible
providing the data span a range of missing energy large enough.

Obviously for the diagonal part of the spectral function 
the well--known sum rule 
\beq
\int \frac{d \vec p}{(2\pi)^3} \int d{\cal E} < \vec p,s 
\vert S \left({\cal E}\right)\vert \vec p,s > = A,
\label{II.18}
\eeq
holds, $A$ being the number of nucleons in the nucleus and a spin--isospin 
summation being implicitly assumed.

Next the matrix elements in momentum space
\beq
< {\vec p}\ ' \vert \hat \rho_\N \vert \vec p > = < {\vec p}\ ' \vert
\widetilde{\vec p}_\N >< \widetilde{\vec p}_\N \vert \vec p > 
\label{II.19}
\eeq
of the {\em one--body projection operator} $\hat \rho_\N$ should be brought 
into the formalism. The operator $\hat \rho_\N$ embodies the distortion 
of the outgoing nucleon's wave and it is clearly of central relevance 
for the present treatment. In PWIA the outgoing nucleon state is a
plane wave and thus (\ref{II.19}) becomes:
\beq
< {\vec p}\ ' \vert \hat \rho_\N \vert \vec p > = 
(2\pi)^6 \delta\left(\pv\,'-\pv_\N\right)\delta\left(\pv_\N-\pv\right)\, .
\label{II.19bis}
\eeq

In the next three sections we shall study in detail both $S$ and $\hat
\rho_\N$ in the semi--classical framework.

Finally the one-body operator associated with the electromagnetic vertices,
namely the {\em single nucleon tensor}
\beq
< {\vec p}\ ', s' \vert W^{\mu\nu}_{sn} \ (\vec q\,) \vert \vec p , s > =
j^{\mu*}_{s_\N\, s'} \ ({\vec p}\ ' + \vec q, {\vec p}\ ') \ j^{\nu}_{s_\N\, s} 
\ (\vec p + \vec q, \vec p)\, ,
\label{II.20}
\eeq
is the third element which enters into the physics of the 
electromagnetic exclusive processes.
Concerning the latter,  we shall assume it to be 
on the mass--shell. Furthermore we shall discuss in the following Section 
the condition under which the off--diagonal matrix elements of 
$W^{\mu\nu}_{sn}$
can be disregarded. The diagonal $W^{\mu\nu}_{sn}$ directly relates 
to the physical ``on shell'' electron--nucleon cross--section according to 
\beq
\left(\frac{d \sigma}{d \Omega_e}\right)_{sn} 
= 2 \alpha^2 \ \frac{1}{Q_{\mu}^4} 
\ \frac{k'}{k} \ \eta_{\mu\nu} 
< \vec p,s \vert W^{\mu\nu}_{sn} (\qv\,)\vert \vec p,s>\, .
\label{II.21}
\eeq

\section{The semi--classical approach}
\setcounter{equation}{0}

For the details of the method we refer the reader to ref.~\cite{REF3}. Here 
it suffices to remind the definition of the Wigner transform (WT) of a 
one--body operator O, namely
\beq
O_W (\vec R, \vec p) = \int \frac{d \vec k}{(2\pi)^3} e^{i \vec k\cdot\vec R}
 \left\langle \vec p + \frac{\vec k}{2} \left| \hat O \right|
\vec p - \frac{\vec k}{2} \right\rangle
\label{III.1}
\eeq
and
\beq
< \vec p\, \vert \, \hat O \ \vert \ {\vec p}\ ' > = \int d \vec R \ e^{i
(\vec p- {\vec p}\ ')\cdot\vec R} 
\ O_W \left(\vec R, \frac{\vec p + {\vec p}\ '}{2}\right).
\label{III.2}
\eeq

The core of the semi--classical approach lies
in the systematic expansion in $\hbar$ (or in the gradient with respect to
$\Rv$ or $\pv\,$) of the Wigner transform of operators; in our case only 
the leading order of the expansion will be kept: accordingly the Wigner
transform of the product of operators reduces to the product of the
Wigner transforms of each factor. We shall repeatedly exploit this rule
in the following.

Let us now start from the exclusive cross--section (\ref{II.10}) which we 
rewrite as follows
\bea
\frac{d^4 \sigma}{d \Omega_e d\epsilon' d \vec p_\N} &&= \frac{2 \alpha^2}
{(2\pi)^3} \ \frac{1}{Q^4_{\mu}} \ \frac{k'}{ k} \eta_{\mu\nu} \int 
\frac{d \vec p}{(2\pi)^3} \int \frac{d {\vec p}\ '}{(2\pi)^3} 
\sum_{s_\N,s',s} < {\vec p\,}' + 
\vec q\,\vert \hat \rho_\N \vert \vec p + \vec q >
\nonumber \\
&&\qquad\times < \vec p, s \vert S ({\cal E}) \vert {\vec p}\ ', s' >
< {\vec p}\ ', s' \vert W^{\mu\nu}_{sn} (\vec q) \vert \vec p , s >.
\label{III.3}
\eea
Performing the change of variables
\beq
\vec p = \vec K + \frac{\vec u}{2} \ \ \ \ {\hbox{and}} \ \ \ \ {\vec p}\ ' =
\vec K - \frac{\vec u}{2}
\label{III.4}
\eeq
and introducing the Wigner transforms, we can recast the above expression 
into the form
\bea
&&\frac{d^4\sigma}{d \Omega_e d\epsilon' d\vec p_\N} \
 = \ \frac{2 \alpha^2}{(2\pi)^3} \ \frac{1}{Q^4_{\mu}} \ \frac{k'}{k} 
\eta_{\mu\nu} \ \int \frac{d \vec K}{(2 \pi)^3} \ \int 
\frac{d \vec u}{(2 \pi)^3} \ \sum_{s_\N s' s}
\nonumber \\
&&\qquad\quad\times
 \int d \vec R\, e^{-i \vec u\cdot \vec R} \ \left[ \hat \rho_\N\right]_W 
\left(\vec R, \vec K + \vec q\, \right) \ \int d \vec S\, 
e^{i\vec u\cdot \vec S} \
\left[ S_{ss'} ({\cal E})\right]_W \left(\vec S, \vec K \right)
\nonumber \\
&&\qquad\quad\times
\left\langle \left.\vec K - \frac{\vec u}{2} , s' \ \right| \ 
W^{\mu\nu}_{sn} (\vec q\,) \ \left| \ \vec K + \frac{\vec u}{2} , s 
\right.\right\rangle
\label{III.5} \\
&&\quad
 = \ \frac{2 \alpha^2}{(2 \pi)^3} \ \frac{1}{Q^4_{\mu}} \ \frac{k'}{k} \
\eta_{\mu\nu} \sum_{s's} 
\ \int  \frac{d\vec K}{(2 \pi)^3} \ \int d \vec R\, d \vec S \
\left[ S_{s's} ({\cal E})\right]_W ( \vec S, \vec K)
\nonumber \\
&&\qquad\quad\times
\left[\hat \rho_\N\right]_W (\vec R, \vec K + \vec q ) \ 
\left[ W^{\mu\nu}_{s', s}(\qv)\right]_W (\vec S - \vec R, \vec K)
\nonumber
\eea
where
\bea
&&\left[ W^{\mu\nu}_{s', s}(\qv\,)\right]_W (\vec S- \vec R , \vec K)\ 
= \ \int \frac{d \vec t}{(2 \pi)^3} \ e^{i\vec t\cdot(\vec S - \vec R)}
\nonumber \\
&&\qquad\quad\times \sum_{s_\N}
\left[ j^{\mu}_{s_\N s'} \left(\vec K + \frac{\vec t}{2} + \vec q\ , 
\vec K + \frac{\vec t}{2}\right)\right]^* \ j^{\nu}_{s_\N s} 
\left(\vec K - \frac{\vec t}{2} + 
\vec q\ , \vec K - \frac{\vec t}{2}\right)
\nonumber \\
&& \qquad\quad\quad
\simeq \delta \left(\vec R - \vec S\,\right) 
< \vec K,s \ \vert \ W^{\mu\nu}_{sn} (\vec q\,) \ \vert \ \vec K, s >.
\label{III.6}
\eea
To leading order in $\hbar$ the $t$ dependence in the electromagnetic 
current can be disregarded. Hence the Wigner transform in eq.(\ref{III.6})
depends only on the diagonal nucleonic matrix elements. This approximation 
leads to a local expression for the exclusive cross-section. 

By inserting (\ref{III.6}) into (\ref{III.5}) one finally obtains for the 
exclusive cross--section the semi--classical expression
\bea
&&\frac{d^4\sigma}{d \Omega_e d \epsilon' d \vec p_\N} = 
\frac{2 \alpha^2}{(2\pi)^3} \ \frac{1}{Q^4_{\mu}} \ \frac{k'}{k} \ 
\eta_{\mu\nu} \int \frac{d \vec K}{(2\pi)^3} 
< \vec K,s \vert \ W^{\mu\nu}_{sn} (\vec q) \vert \vec K,s>
\nonumber \\
&&\qquad\quad \times  \int d \vec R \left[ S_{ss} ({\cal E})\right]_W
(\vec R , \vec K) \left[\hat \rho_\N\right]_W (\vec R, \vec K +\vec q)
\label{III.7} \\
&&\quad =  \frac{1}{(2\pi)^3}\,\int \frac{d \vec K}{(2\pi)^3}\,
\left(\frac{d \sigma}{d \Omega_e}\right)_{sn}(\vec K, \vec q)
\,\int d \vec R \,\left[ S_{ss}({\cal E})\right]_W (\vec R, \vec K)
\, \left[\hat \rho_\N\right]_W (\vec R, \vec K + \vec q) 
\nonumber
\eea
(repeated indices are meant to be summed) where
\beq
\left[\frac{d\sigma}{d\Omega_e}\right]_{sn}\!(\pv, \qv) =
\sigma_{Mott}\frac{M_\N}{E(p)}\frac{M_\N}{E(|\pv+\qv\,|)} v_L R_L
\label{VII.10}
\eeq
$\sigma_{Mott}$ being the Mott cross--section, $E(p)=\sqrt{p^2+M_\N^2}$,
\beq
v_L=\left(\frac{Q^2_\mu}{q^2}\right)^2\equiv 
\left(\frac{\tau}{\kappa^2}\right)^2\, , 
\label{VII.11a}
\eeq
\beq
 R_L=\frac{\kappa^2}{\tau}\left\{G_E^2(\tau) + 
W_2(\tau)\chi^2\right\}
\label{VII.11b}
\eeq
and
\beq
W_2(\tau)= \frac{1}{1+\tau}\left[G_E^2(\tau) + \tau G_M^2(\tau)
\right]\, . 
\label{VII.11c}
\eeq
Furthermore the dimensionless variables $\chi=p\sin\theta/M_\N$ ($\theta$
being the angle between $\pv$ and $\qv$), $\kappa=q/2M_\N$, $\lambda=
\omega/2M_\N$, $\tau=\kappa^2-\lambda^2$ are utilized, and $G_E$ and $G_M$
are the Sach's electric and magnetic nucleon's form factors (only the 
longitudinal part of the cross--section is considered here, for the sake of 
illustration).

Formula (\ref{III.7})  clearly shows how 
the PWIA scheme has been improved. Indeed, in the framework of the PWIA, 
only one nucleon inside the nucleus, with a given momentum $\vec K$, 
takes part in the exclusive process leading to a final
nucleon with momentum $\vec p_\N = \vec K + \vec q$, whereas now, 
because of the FSI embodied in ${\hat\rho}_\N$, 
the momentum of the nucleon actually involved into the process might take
any value: hence the integration over the variable $\vec K$.

The above formulae also transparently show how the Wigner transform
operates: it replaces the off--diagonal momentum matrix elements of the
one--body operators entering into the expression for the exclusive 
cross--section with diagonal ``space-dependent'' matrix elements. 
Moreover the WT of the matrix elements of the one--body operators 
associated with the distortion of the outgoing nucleon wave, the spectral 
function and the single nucleon cross--section are each evaluated at 
``different places'' into the nucleus and 
finally their product is integrated over the whole nuclear volume.

This statement is rigorously true when one does not employ the 
diagonal approximation (\ref{III.6}). When the latter is adopted,
however, then the various ingredients of the exclusive 
cross--section are actually evaluated at the same place. 
It thus appears that the exclusive process, in the semi--classical 
framework, is viewed as being built up ``non locally'' 
(but ``locally'' in the present,
leading order approximation) from elementary contributions arising
from different regions of the nucleus. 
The interference among these is neglected
in first order, but it can be accounted for through higher order terms in the
$\hbar$ expansion.

Before ending this section we write the expression for the semi--classical
PWIA cross--section; introducing into (\ref{III.3}) 
the explicit form (\ref{II.19bis}) for the
matrix element of the one--body projection operator 
and applying again the WT to the spectral function and the single nucleon 
tensor, one easily ends up with:
\beq
\left(\frac{d^4\sigma}{d \Omega_e d \epsilon' d \vec p_\N}\right)_{PWIA} 
= \frac{1}{(2\pi)^3}\, 
\left(\frac{d \sigma}{d \Omega_e}\right)_{sn}(\pv_m, \vec q)
\,\int d \vec R \,\left[ S_{ss}({\cal E})\right]_W (\vec R, \pv_m)
\label{pwia}
\eeq
where the so--called missing momentum $\pv_m=\pv_\N-\qv$ has been 
introduced. The above formula clearly shows that in PWIA the cross--section
is directly proportional to the spectral function.

In the next two sections we shall calculate the WT of the spectral function and
of the distortion operator in leading order.

\section{ The diagonal semi--classical spectral function}
\setcounter{equation}{0}

Let us consider a closed shell nucleus having $A$ nucleons sitting in the lowest
orbits of a potential well V(R) (in principle the Hartree--Fock (HF) mean 
field) and an $(A-1)$ daughter nucleus obtained by creating a hole 
in a generic occupied level of the former. 
The $(A-1)$ nucleus thus obtained will generally be in an excited
state with an energy given by (neglecting the small recoil energy)
\bea
&&E_{A-1} \equiv E_{A-1,h} 
\label{IV.1} \\
&&= \sum_{\beta < \epsilon_F} \ t_\beta + \frac{1}{2} \ 
\sum_{\beta,\beta'} < \beta \beta' \vert v \vert \beta\beta' >_a + AM_\N
-t_h - \sum_\beta < h \beta \vert v \vert h \beta >_a - M_\N
\nonumber 
\eea
in the non--relativistic HF approximation. In (\ref{IV.1}) the first 
three terms 
on the RHS correspond to the energy of the nucleus with mass number $A$, 
$v$ is a suitable two--body interaction and the subtracted terms represent 
the energy of a particle in the $h$ orbit. Accordingly we can recast 
(\ref{IV.1}) as follows 
\beq
E_{A-1,h}=M_A-M_\N-\epsilon_h\, ,
\label{IV.1a}
\eeq
which defines the hole energy $\epsilon_h$. Of course in the HF scheme
the ground state energy of the $A-1$ daughter nucleus is obtained by 
removing a particle at the Fermi energy, namely 
\beq
E_{A-1}^0 = M_A- M_\N -\epsilon_F\equiv M_{A-1}\, .
\label{IV.1b}
\eeq
from where the relation $\epsilon_F= -E_S$ follows. We 
then get for the (positive) excitation energy 
\beq
\cale = E_{A-1,h}-M_{A-1}=\epsilon_F-\epsilon_h\, .
\label{IV.1c}
\eeq
In the semi--classical approximation  the ``negative'' Fermi energy is
\beq
\epsilon_F = \frac{k_F^2 (R)}{2 M^*_\N} + V(R)
\label{IV.2}
\eeq
$R$ being the radial variable; a nucleon effective mass $M^*_\N$ has
been introduced to account, together with the potential well V(R), for the
Hartree--Fock mean field. Equation (\ref{IV.2}) locally defines 
a Fermi momentum $k_F(R)$ for $R\le R_c$, $R_c$ being the classical
turning point fixed by the equation 
\beq
\epsilon_F=V(R_c)\, . 
\label{IV.2a}
\eeq

We now express the spectral function in the basis of the eigenfunctions 
of the single particle hamiltonian
\beq
h = {p^2 \over 2 M^*_\N} + V(R)\, ,
\label{IV.6}
\eeq
rather than in the basis of the momentum eigenfunctions. The former
of course obey the equation
\beq
h \vert \alpha > = \epsilon_\alpha \vert \alpha >
\label{IV.7}
\eeq
$\epsilon_\alpha$ being the associated eigenvalues.

We thus get 
\bea
&&< {\vec p}\  \vert \ S({\cal E}) \ \vert \ {\vec p}\ ' >
= \sum_{\alpha\alpha'} < \vec p \ '\vert \alpha' > < A \ \vert \ \hat
a^\dagger_{\alpha'}\, \delta \left(\widehat{\cal H} - {\cal E}\right)
\, \hat a_{\alpha} \ \vert A > < \alpha \ \vert \ {\vec p}\ >
\nonumber \\
&&\qquad = 
\sum_{\alpha\alpha'} < \vec p \ '\vert \ \alpha' > < \alpha' \ \vert \
\delta \left\lbrack {\cal E} - (\epsilon_F - \epsilon_\alpha)\right\rbrack 
\theta(\epsilon_F-\epsilon_\alpha)
\vert \ \alpha > < \alpha \ \vert \ {\vec p}\  >
\nonumber \\
&&\qquad\quad =< \vec p \ '\vert \ \theta (\epsilon_F - h) \ 
\delta \left\lbrack {\cal E} - (\epsilon_F - h) \right\rbrack 
\ \vert \ {\vec p}\  >\, .
\label{IV.8}
\eea

By applying then the definition (\ref{III.1}) of the Wigner transform 
it is an easy matter to verify that, in the leading order of the 
$\hbar$ expansion, one has
\beq
\left\lbrack \theta (\epsilon_F - h)\right\rbrack_W (\vec R,\vec p) = 
\theta \left\lbrack \epsilon_F - \left( \frac{p^2}{2 M^*_\N} + 
V(R)\right) \right\rbrack + O(\hbar^2)
\label{IV.9a}
\eeq
and
\beq
\left\lbrace\delta \left\lbrack {\cal E} - (\epsilon_F - h)
\right\rbrack\right\rbrace_W (\vec R,\vec p) = \delta \left\lbrack {\cal E} - 
\left(\epsilon_F - {p^2 \over 2 M^*_\N} - V(R)
\right) \right\rbrack + O(\hbar^2)\, .
\label{IV.9b}
\eeq
Hence it follows that, in leading order, the WT of the spectral function is
just the product of the WT (\ref{IV.9a}) and (\ref{IV.9b}), namely
\beq
\left\lbrack S ({\cal E})\right\rbrack_W (\vec R, \vec p) = 
\theta \left( \epsilon_F - \frac{p^2}{2 M^*_\N} - V(R)\right) 
\ \delta \,\left\lbrack {\cal E} - \left(\epsilon_F - 
\frac{p^2}{2 M^*_\N} - V(R)\right) \right\rbrack
\label{IV.10}
\eeq
or, recalling (\ref{IV.2}),
\beq
\left\lbrack S ({\cal E})\right\rbrack_W (\vec R, \vec p) = 
\theta \left( \frac{k^2_F (R)}{ 2 M^*_\N} - \frac{p^2}{2 M^*_\N}\right) 
\ \delta \,\left\lbrack {\cal E} - \left( \frac{k^2_F (R)}{2 M^*_\N} - 
\frac{p^2}{2 M^*_\N}\right) \right\rbrack\, .
\label{IV.11}
\eeq

An integration over the whole nucleus yields then for the diagonal spectral
function in the semi--classical approximation the expression
\beq
S (\vec p,{\cal E}) = \int d \vec R \, 
\theta \left(\frac{k_F^2 (R)}{2 M^*_\N} - \frac{p^2}{2 M^*_\N}\right) 
\ \delta \,\left\lbrack {\cal E} - \left( \frac{k^2_F(R)}{2 M^*_\N} - 
\frac{p^2}{2 M^*_\N} \right) \right\rbrack
\label{IV.12}
\eeq
which displays a striking similarity with the one of a Fermi gas.
Actually, in leading order, the HF semi--classical 
approximation of the spectral
function for a finite nucleus might be viewed as arising from a superposition
of a large set of FG each one characterized by a different $k_F$. 
The latter is locally defined according to the prescription
\beq
k_F (R) = \sqrt{2 M^*_\N \left(\epsilon_F - V(R)\right)}\, . 
\label{IV.13}
\eeq
Note that, as previously mentioned, there is no interference among the
contributions of the different volume elements to the spectral function in
leading order.

To illustrate  the method we report in the Appendix the calculation of
the diagonal semi--classical spectral function for a few 
specific single particle potentials.

\section{The distortion operator}
\setcounter{equation}{0}

The distortion operator
\beq
\widehat{\rho}_\N = \ \vert \ \widetilde{\vec p}_\N >< \widetilde{\vec p}_\N \
\vert
\label{V.1}
\eeq
obeys, in the HF scheme, the equation
\beq
h \ \widehat{\rho}_\N = \frac{p^2_\N}{2 M_\N} \ \widehat{\rho}_\N
\label{V.2}
\eeq
where, on the RHS, the free nucleon mass appears.

Taking the WT of the above one gets, in first order,
\beq
(h \widehat{\rho}_\N)_W \ \cong \ (h)_W \ (\widehat{\rho}_\N)_W \ = 
\ \frac{p^2_\N}{2 M_\N} \ (\widehat{\rho}_\N)_W
\label{V.3}
\eeq
or, focussing on the dependence upon the variables $\vec R$ and $\vec p$,
\beq
\left\lbrack (h)_W \ (\vec R, \vec p) - \frac{p^2_\N}{2 M_\N}\right\rbrack \
(\widehat{\rho}_\N)_W \ (\vec R, \vec p) = 0\, .
\label{V.4}
\eeq
Now since
\beq
(h)_W \ (\vec R, \vec p) \ = \ \frac{p^2}{2 M^*_\N} \ + \ V(R) ,
\label{V.5}
\eeq
V(R) being the chosen potential well, one can set
\beq
(\widehat{\rho}_\N)_W \ (\vec R, \vec p) \ = 
\ Z \ (\vec R, \vec p\, ; \vec p_\N) \
\delta \ \left(\frac{p^2_\N}{2 M_\N} - V(R) \ - \ \frac{p^2}{2M^*_\N}\right)
\label{V.6}
\eeq
where the factor $Z (\vec R, \vec p \,  ; \vec p_\N)$ can be partially fixed 
by closure.

The latter indeed requires
\beq
\int \frac{d \vec p_\N}{(2\pi)^3} \ \vert \ \widetilde{\vec p}_\N ><
\widetilde{\vec p}_\N \ \vert \ + \sum_{\alpha} \ \vert \ \alpha >< \alpha \
\vert \,\,\theta( - h) = 1
\label{V.7}
\eeq
the sum being extended over the bound HF orbits.
In Wigner transform (\ref{V.7}) becomes
\bea
\int \frac{d \vec p_\N}{(2\pi)^3} \ {\left(\widehat{\rho}_\N\right)}_W 
(\vec R, \vec p) &=&
\int \frac{d \vec p_\N}{(2 \pi)^3} Z (\vec R, \vec p\, ; \vec p_\N) \ \delta \
\left(\frac{p^2_\N}{2 M_\N} - V(R) - \frac{p^2}{2 M^*_\N}\right) 
\nonumber \\
&=& \theta \left(\frac{p^2}{2 M^*_\N} + V(R)\right)
\label{V.8}
\eea
or, after performing the integration over the modulus of $\vec p_\N$,
\beq
M_\N p_\N (R,p) \int 
\frac{d \widehat{p}_\N}{(2\pi)^3} \ Z \Big(\vec R, \vec p \, 
; \vec p_\N(R,p)\Big) = \theta \left(\frac{p^2}{2 M^*_\N} + V(R)\right)
\label{V.9}
\eeq
the integral being over the direction of $\vec p_\N(R,p) \equiv p_\N (R,p) 
\widehat{p}_\N$, with
\beq
p_\N (R,p) = \sqrt{\frac{M_\N}{ M^*_\N}\, p^2 + 2 M_\N V(R)}\, .
\label{V.10}
\eeq

In a quantum framework one would obtain the states 
$|{\widetilde{\vec p}_\N}>$ from the positive energy solutions of an 
appropriate optical potential: the distortion operator, as well as its
Wigner transform (\ref{V.6}), would then be fixed.  
Here, in the spirit of the semi--classical 
approach, we heuristically set 
\beq
Z \left(\vec R, \vec p ; \vec p_\N (R,p) \right) = 
\frac{(2\pi)^3}{M_\N p_\N (R,p)} \ F(\widehat{p}_\N, \widehat{p})\,\, 
\theta \left(\frac{p^2}{2 M^*_\N} + V(R)\right)\, ,
\label{V.11}
\eeq
with the additional condition
\beq
\int d \widehat{p}_\N \ F(\widehat{p}_\N, \widehat{p}) = 1\, .
\label{V.12}
\eeq

As an illustration we shall consider in the following two extreme 
assumptions for $F$, namely:
\begin{itemize}
\item
\beq
F (\widehat{p}_\N, \widehat{p}) = \delta \left(\widehat{p}_\N -
\widehat{p}\right)
\label{V.13}
\eeq
which corresponds to the {\em eikonal approximation} and therefore is 
expected to be valid only for large enough energies of the outgoing nucleon. 
In this case in fact the final nucleon should be little deflected from the
direction of the initial one, which has absorbed the photon inside 
the nucleus;
\item
\beq
F(\widehat{p}_\N, \widehat{p}) = {1 \over 4\pi}
\label{V.14}
\eeq
which obviously corresponds to a final nucleon escaping the
nucleus with the same probability in all directions ({\em uniform 
approximation}).
\end{itemize}

\noindent
The true physics should of course lie in between the predictions of 
(\ref{V.13}) and (\ref{V.14}), respectively.

\section{The semi--classical exclusive cross--section}
\setcounter{equation}{0}

In this Section we derive explicit expressions for the exclusive 
cross--section in the
semi--classical approximation, treating the distortion of the outgoing 
nucleon wave in the HF approximation as discussed in Section 5. We use
the harmonic oscillator 
(cut at the classical turning point) and the Woods--Saxon wells. 

To settle the basis for this scope, we insert into the expression of the
exclusive cross--section (\ref{III.7}) the distortion operator as 
given by (\ref{V.6}) and (\ref{V.11}). Then the $\delta$--function appearing 
into the latter allows us to perform the integration over the modulus 
of the momentum variable. We thus obtain:
\bea
 \frac{d^4\sigma}{d\Omega_e d\epsilon' d\pv_\N}&& =
\frac{1}{(2\pi)^3}\left(\frac{M^*_\N}{M_\N}\right) \int d{\hat p}\,
\int d\Rv \left[{d\sigma\over {d\Omega_e}}\right]_{sn}\!
\left({\PR}{\hat p} - \qv, \qv\right) 
\nonumber \\
&& \times\left[S_{ss}({\cal E})\right]_W(\Rv, {\PR}{\hat p} -\qv)
{\PR\over p_\N}F({\hat p_\N},{\hat p})
\label{VI.1}
\eea
with
\beq
{\PR}=\sqrt{{M_\N^*\over M_\N}}\sqrt{{p_\N^2}-2 M_\N V(R)}\, .
\label{VI.2}
\eeq

To proceed further we exploit the general expression (\ref{IV.10}) for the
semi--classical spectral function in Wigner transform. Then (\ref{VI.1}) 
can be recast as follows
\bea
&&\frac{d^4\sigma}{d\Omega_e d\epsilon' d\pv_\N} =
\frac{1}{2\pi^2}\left(\frac{M^*_\N}{M_\N}\right) \int R^2 dR \ 
\frac{\PR}{|\qv-\pv_m|}
\int d{\hat p}\, F\left({\widehat{\qv-\pv_m}},{\hat p}\right)
\left[\frac{d\sigma}{d\Omega_e}\right]_{sn}\!
\left({\PR}{\hat p} - \qv, \qv\right) 
\nonumber\\
&&\qquad\qquad \times \delta \left[{\cal E} -
\left(\epsilon_F - \frac{(\qv-\pv_m)^2}{2M_\N} - 
\frac{q^2}{2M^*_\N} + \frac{\PR}{M^*_\N}{\hat p}\cdot\qv
\right)\right]\theta(\cale)
\label{VI.3} 
\eea
where the trivial integration over the angles of the vector $\Rv$ 
has been performed and the ``missing momentum'' variable 
$\pv_m=\qv-\pv_\N$ has been used.

We now separately investigate the exclusive cross section in the two
limiting ap\-pro\-xi\-mations for the ``distortion function'' $F$ 
discussed at the end of Section 5 (for simplicity here and in the 
following we shall ignore the effective mass, setting $M^*_\N=M_\N$). 

To start with we consider the {\it eikonal approximation} [formula 
(\ref{V.13})]. In this case it is straightforward to obtain the following 
one--dimensional integral expression for the exclusive cross--section
\bea
&& \frac{d^4\sigma}{d\Omega_e d\epsilon' d\pv_\N} =
\frac{1}{2\pi^2} \int_0^{R_{c}} R^2 dR
\frac{\PR}{|\qv-\pv_m|}
\left[\frac{d\sigma}{d\Omega_e}\right]_{sn}\!
\left(\frac{\PR(\qv-\pv_m)}{|\qv-\pv_m|}-\qv,\qv\right)
\nonumber \\
&&\quad \times \delta \left\{ {\cal E} -
\left[\epsilon_F - \frac{p^2_m}{2M_\N}
+\left(\frac{q^2-\qv\cdot\pv_m}{M_\N}\right)
\left(\frac{\PR}{|\qv-\pv_m|}-1\right)\right]\right\}\theta(\cale)\, .
\label{VI.4}
\eea
Note that the ``exclusive variables'' ${\cal E}$ and $p_m$ appear explicitly
in (\ref{VI.4}); among the ``inclusive'' ones only 
$q$ does, whereas the transferred
energy $\omega$ is hidden in the scalar product ${\vec q}\cdot\pv_m$.
Moreover the upper limit of the $R$--integration in (\ref{VI.4}) is set 
by the equation (\ref{IV.2a}),
which clearly entails $V(R_{c})<0$ and 
$\sqrt{1-2M_\N V(R)/(\qv-\pv_m)^2} >1 $.
Hence, since the hole energy
\beq
\epsilon_h= \frac{p^2_m}{2M_\N} - \frac{q^2-\qv\cdot\pv_m}{M_\N}
\left(\frac{\PR}{|\qv-\pv_m|}- 1\right)
\label{VI.6}
\eeq
is obviously negative, it follows that in the semi--classical eikonal
approximation the projection of the missing momentum $\pv_m$ on the momentum 
transfer $\qv$ has to be less than $q$.

It remains now to exploit the energy conserving $\delta$--function in 
(\ref{VI.4}) to perform the $R$--integration. This is easily achieved 
by taking advantage of the identity 
\beq
\delta\left[{\cal E}-\left(\epsilon_F-\epsilon_h\right)\right]=
\frac{\left(\qv-\pv_m\right)^2}{\left(q^2 - \qv\cdot\pv_m\right)}\,
\frac{\delta(R-{\tilde R})}{\left|dV/dR\right|}\,
\frac{\PR}{|\qv-\pv_m|}\, , 
\label{VI.7}
\eeq
${\tilde R}$ being the root of the equation
\beq
{\cal A}(\tilde R) = {\cal A} \equiv
\frac{ {\cal E}-\epsilon_F +{p^2_m/{2M_\N}}}
{\left(q^2-\qv\cdot\pv_m\right)/M_\N} + 1 \, 
\label{VI.8}
\eeq
where
\beq
{\cal A}(R)\equiv \frac{\PR}{|\qv-\pv_m|}
= \sqrt{1-\frac{V(R)}{(\qv-\pv_m)^2/2M_\N}}\, .
\label{VI.9}
\eeq
One then obtains the following ``analytic'' expression for the 
semi--classical exclusive cross--section in the eikonal approximation 
for the mean field distortion operator:
\beq
\frac{d^4\sigma}{d\Omega_e d\epsilon' d\pv_\N} =
\frac{1}{2\pi^2} {\tilde R}^2 {\cal A}^2({\tilde R})
\frac{1}{\left|dV/dR\right|_{R={\tilde R}}}\,
\frac{\left(\qv-\pv_m\right)^2}{\left(q^2 - \qv\cdot\pv_m\right)}
\left[\frac{d\sigma}{d\Omega_e}\right]_{sn}\!
\left[{\cal A}(\qv-\pv_m) - \qv, \qv\right]\, .
\label{VI.10}
\eeq

For a full exploitation of formula (\ref{VI.10}) an expression for the 
cosine of the angle between $\pv_m$ and $\qv$ is still needed: it is most 
easily obtained by applying the energy conservation to the right sector 
of the diagram displayed in Fig.~2. One gets:
\beq
\cos\theta_{p_m q}=
\frac{M_\N}{ qp_m}\left({\cal E}+E_S-\omega\right)
+\frac{1}{2}\left[ \frac{p_m}{q}\left(1+\frac{M_\N}{M_{A-1}}\right)
+ \frac{q}{p_m}\right]\, .
\label{VI.11}
\eeq

Choosing now for $V(R)$ the harmonic oscillator
potential, as given by formula (\ref{IV.19}) with the bare nucleon's mass,
we easily get the following cross--section
\bea
&&\frac{d^4\sigma}{d\Omega_e d\epsilon' d\pv_\N} =
\frac{1}{\sqrt{2}\pi^2} \frac{1}{M_\N^{3/2}\omega_o^3}
\left\{ \frac{(\qv-\pv_m)^2}{2M_\N}\left(1-{\cal A}^2\right)
+ V_o\right\}^{1/2}
\nonumber \\
&&\quad\qquad \times {\cal A}^2 \frac{\left(\qv-\pv_m\right)^2}
{\left(q^2 - \qv\cdot\pv_m\right)}
\left[\frac{d\sigma}{d\Omega_e}\right]_{sn}\!
\left[{\cal A}(\qv-\pv_m) - \qv, \qv\,\right]
\label{VI.12}
\eea
where the solution of (\ref{VI.8}) is 
\beq
{\widetilde R}=\frac{1}{\omega_o}\sqrt{ \frac{2}{M_\N}
\left[ \frac{(\qv-\pv_m)^2}{2M_\N}\left(1-{\cal A}^2\right) +V_o
\right]} \, .
\label{VI.13}
\eeq

For a Woods--Saxon well 
\beq
V(R) = - \frac{V_1}{1+e^{(R-R_o)/a}}\, ,
\label{VI.13a}
\eeq
formula (\ref{VI.10}) holds  with
\beq
{\widetilde R}= R_o + a\ln\left\{ 
\frac{2M_\N V_1}{(\qv-\pv_m)^2\left({\cal A}^2-1\right)} 
-1 \right\}\, .
\label{VI.13b}
\eeq

Next we turn to the {\it uniform approximation} for the distortion
function $F$ [formula (\ref{V.14})]. In this case 
a fully analytic expression for the exclusive cross--section cannot
be achieved. Indeed by exploiting the $\delta$--function of the 
distortion operator and the azimuthal angle independence of the 
elementary single nucleon cross--section, one gets
\bea
&&\frac{d^4\sigma}{d\Omega_e d\epsilon' d\pv_\N} =
\frac{1}{(2\pi)^2} \int_0^{R_{c}} R^2 dR\, \frac{\PR}{|\qv-\pv_m|}
\int d\cos\theta 
\left[\frac{d\sigma}{d\Omega_e}\right]_{sn}\!\left({\PR}{\hat p}-\qv,
\qv\,\right)
\nonumber \\
&&\qquad\quad\times 
\delta \left\{ {\cal E} -
\left[\epsilon_F - \frac{(\qv-\pv_m)^2}{2M_\N}-\frac{q^2}{2M_\N}
+\frac{\PR}{M_\N}\,q\cos\theta\right]\right\}
\label{VI.14} 
\eea
where ${\PR}$ and $R_{c}$ are again fixed by eq.(\ref{VI.2}) and
(\ref{IV.2a}), and $\theta$ is the angle between $\qv$ and $\pv$. 
The integration over the latter variable is trivial and one
finally obtains for the semi--classical exclusive cross--section, in the
uniform approximation for the distortion operator, the following 
one--dimensional integral expression
\bea
&&\frac{d^4\sigma}{d\Omega_e d\epsilon' d\pv_\N} =
\frac{1}{(2\pi)^2} \frac{M_\N}{q}\frac{1}{|\qv-\pv_m|}
\int_0^{R_{c}} R^2 dR 
\label{VI.15} \\
&&\qquad\quad\times \left.
\left[ \frac{d\sigma}{d\Omega_e}\right]_{sn}\!\left({\PR}{\hat p}-\qv,
\qv\, \right) \right|_{\cos\theta=y_0(R)} \theta(1-|y_0(R)|)
\nonumber
\eea
where
\beq
y_0(R)= \frac{M_\N}{{\PR}q}\left[ {\cal E} -
\left(\epsilon_F - \frac{(\qv-\pv_m)^2}{2M_\N}-\frac{q^2}{2M_\N}\right)
\right]\, .
\label{VI.16}
\eeq

Notably the one--body potential confining the nucleons into the 
nucleus does not explicitly appear in (\ref{VI.15}): 
it is however hidden in the equations fixing $R_{c}$, $\epsilon_F$
and $\PR$.

\section{The semi--classical $t$--inclusive cross--section}
\setcounter{equation}{0}

To get the inclusive cross--section in the $t$--channel one should 
integrate the exclusive one, obtained in the previous Section, over
the outgoing nucleon's momentum. Indeed in the $t$--inclusive scattering 
only the final electron is detected: accordingly the momentum $\qv$ 
transferred to the nucleus is kept fixed in the process.
By contrast in the $u$--channel, where the outgoing nucleon only is 
detected, the vector ${\vec\xi}=\pv_\N-\kv$ is kept fixed, whereas $\qv$
varies.

We shall apply the semi--classical formalism to the $u$--inclusive 
scattering in a forthcoming paper: here we focuss instead on the
$t$--inclusive channel, where the vast majority of the electron
scattering experiments have been performed. 

To start with we first show that by integrating the exclusive cross--section
over the momentum of the emitted nucleon we recover the correct inclusive
cross--section if and only if the distortion of the outgoing particle is
properly accounted for. For this purpose the integral of (\ref{III.7}) 
over $\pv_\N$, 
\bea
\frac{d^2\sigma}{d\Omega_e d\epsilon'}&&=
\int\frac{d\pv_\N}{(2\pi)^3}\int \frac{d{\vec K}}{(2\pi)^3}
\left( \frac{d\sigma}{d\Omega_e}\right)_{sn}\!\left({\vec K}, \qv\right)
\nonumber \\
&&\times \int d\Rv \left[S_{ss}(\cale)\right]_W\left(\Rv,{\vec K}\right)
\left({\widehat\rho}_\N\right)_W\left(\Rv,{\vec K}+\qv\right),
\label{VII.0}
\eea
is carried out by exploiting the semi--classical expressions (\ref{IV.11}) 
for the spectral function and (\ref{IV.2}) for the Fermi energy. We get
\bea
\frac{d^2\sigma}{d\Omega_e d\epsilon'}&&=
\int\frac{d\pv_\N}{(2\pi)^3}\int \frac{d{\vec K}}{(2\pi)^3}
\left( \frac{d\sigma}{d\Omega_e}\right)_{sn}\!\left({\vec K}, \qv\,\right)
\int d\Rv {\left({\widehat\rho}_\N\right)}_W\left(\Rv,{\vec K}+\qv\right)
\nonumber \\
&&\times \theta\left[\epsilon_F-\frac{K^2}{2M^*_\N}-V(R)\right]
\delta\left\{\cale-\left[\epsilon_F-\frac{K^2}{2M^*_\N}-V(R)\right]
\right\}
\label{VII.0a}
\eea
Now, by taking advantage of the $\delta$ explicitly embodied in the 
distortion operator [see (\ref{V.6})] and since (\ref{II.15}) implies 
(neglecting the recoil energy)
\beq
\cale=
\omega +\epsilon_F-V(R)-{(\qv+\Kv)^2\over{2M^*_\N}}\, ,
\label{VII.0b}
\eeq
we obtain
\bea
&&\frac{d^2\sigma}{d\Omega_e d\epsilon'}=
\int\frac{d\pv_\N}{(2\pi)^3}\int \frac{d{\vec K}}{(2\pi)^3}
\left( \frac{d\sigma}{d\Omega_e}\right)_{sn}\!\left({\vec K}, \qv\right)
\int d\Rv \left({\widehat\rho}_\N\right)_W\left(\Rv,{\vec K}+\qv\right)
\nonumber \\
&&\qquad\times 
\theta\left[\epsilon_F-\frac{K^2}{2M^*_\N}-V(R)\right]
\delta\left\{\omega-\left(\frac{q^2}{2M^*_\N}+\frac{\qv\cdot\Kv}{M^*_\N}
\right)\right\}\, .
\label{VII.0c}
\eea
Then the integration over $\pv_\N$ is immediately done with the help of
(\ref{V.8}) and we end up with
\bea
&& \frac{d^2\sigma}{d\Omega_e d\epsilon'}=
\int \frac{d{\vec K}}{(2\pi)^3}\int d\Rv 
\left( \frac{d\sigma}{d\Omega_e}\right)_{sn}\!\left({\vec K}, \qv\right)
\theta\left[\frac{|\Kv+\qv|^2}{2M^*_\N} +V(R)\right]
\nonumber \\
&&\qquad\times 
\theta\left[\epsilon_F-\frac{K^2}{2M^*_\N}-V(R)\right]
\delta\left\{\omega-\left(\frac{q^2}{2M^*_\N}+ 
\frac{\qv\cdot\Kv}{M^*_\N}
\right)\right\}\, ,
\label{VII.0d}
\eea
which is the well--known semi--classical expression for the inclusive
cross--section, limited however to the continuum spectrum for the
emitted nucleon. The internal consistency of the semi--classical 
approach is thus proved.
 In connection with this result we note that it extends
the PWIA of ref.\cite{REF7} where it is shown that for the fully 
quantum mechanical
relativistic Fermi gas the integral of the exclusive cross--section leads to 
the inclusive one only in the non--Pauli blocked domain.\footnote{
The same result is obtained in the present framework, when the PWIA 
for ${\hat\rho}_\N$, expression (\ref{II.19bis}), 
is employed; from the latter, indeed, one gets:
\bea
&& \frac{d^2\sigma}{d\Omega_e d\epsilon'}=
\int \frac{d{\vec K}}{(2\pi)^3}\int d\Rv 
\left( \frac{d\sigma}{d\Omega_e}\right)_{sn}\!\left({\vec K}, \qv\,\right)
\nonumber \\
&&\qquad\times 
\theta\left[\epsilon_F-\frac{K^2}{2M^*_\N}-V(R)\right]
\delta\left\{\omega-\left(\frac{q^2}{2M^*_\N}+ 
\frac{\qv\cdot\Kv}{M^*_\N}
\right)\right\}\, .
\nonumber
\eea} 
The expression (\ref{VII.0d}) instead fully respects the Pauli 
principle: of course it does not account 
for the contribution to the inclusive cross section arising from the
unoccupied bound states lying in between the Fermi energy and the continuum.

For the actual evaluation of (\ref{VII.0d}), however, we choose here to 
follow the approach of
integrating over the ``exclusive'' variables ${\cale}$ and $p_m$. 
For this purpose we observe that, owing to the independence of the 
exclusive cross--section upon the azimuthal angle of $\pv_\N$, the 
integration over the latter can be converted into an integration over 
an appropriate domain of the missing energy -- missing momentum plane. 
Indeed the following relationship holds:
\beq
 \frac{d^2\sigma}{d\Omega_e d\epsilon' } =
2\pi \frac{M_\N}{q}\int_{\Gamma} p_m dp_m\, d{\cal E} 
\frac{d^4\sigma}{d\Omega_e d\epsilon' d\pv_\N}\, ,
\label{VII.1}
\eeq
the boundaries of the integration domain $\Gamma$ being given (in the
positive quadrant of the $(\cale ,p_m)$ plane) by the curves
\bea
\cale^- &=& \omega - E_S -\frac{(q-p_m)^2}{2M_\N} -
\frac{p^2_m}{2M_{\scriptscriptstyle{A-1}}}
\label{VII.2a} \\
\cale^+ &=& \omega - E_S -\frac{(q+p_m)^2}{2M_\N} -
\frac{p^2_m}{2M_{\scriptscriptstyle{A-1}}},
\label{VII.2b}
\eea
which are obtained by setting $\cos\theta_{{p_m}q}=-1$ and 
$\cos\theta_{{p_m}q}=+1$, respectively, in equation (\ref{VI.11}).
They represent the larger ($\cale^-$) and the lower ($\cale^+$)
excitation energy of the residual nucleus compatible with the
kinematical constraints in an exclusive process.

We remind that, while $\cale^-$ always extends to the first quadrant 
of the $(\cale, p_m)$ plane, this might not be the case for $\cale^+$.
Indeed, for $\omega\le  q^2/2M_\N +E_S\equiv 
{\widetilde\omega}$, $\cale^+$ lives entirely outside the first 
quadrant: in this case the lower limit of integration over the variable
$\cale$ should simply be set equal to zero. On the other hand, for
$\omega\ge {\widetilde\omega}$, $\cale^+$ extends to the first
quadrant as well.

Let's first consider the eikonal approximation for the distortion 
function, with the harmonic oscillator as a binding  potential. 
In this instance we obtain for the semi--classical inclusive cross--section:
\bea
&& \frac{d^2\sigma}{d\Omega_e d\epsilon' } =
\sqrt{\frac{2}{M_\N}}\frac{1}{\pi q\omega_o^3} 
\left\{ \theta\left(\omega-{\widetilde\omega}\right)
\left[\int_0^{p^+_{m,max}} p_m dp_m\int_{\cale^+}^{\cale^-} d\cale
+ \int_{p^+_{m,max}}^{p^-_{m,max}} p_m dp_m \int_0^{\cale^-}d\cale
\right]\right. 
\nonumber \\
&&\qquad\qquad \left. + \theta\left({\widetilde\omega}-\omega\right)
\int_{p^-_{m,min}}^{p^-_{m,max}} p_m dp_m\int_0^{\cale^-} d\cale
\right\} 
\nonumber \\
&&\qquad \times
\left\{ \frac{(\qv-\pv_m)^2}{2M_\N}\left[ 1- {\cal A}^2(\cale,p_m)
\right] + V_o\right\}^{1/2} {\cal A}^2(\cale,p_m)
\nonumber \\
&&\qquad\times
\frac{(\qv -\pv_m)^2}{(q^2-\qv\cdot\pv_m)}
\left[\frac{d\sigma}{d\Omega_e}\right]_{sn}\!\left[
{\cal A}(\cale,p_m)(\qv-\pv_m) -\qv, \qv\right] 
\label{VII.3} 
\eea
where one should recall that ${\cal A}$, defined in eq.~(\ref{VI.8}),
explicitly depends upon $\cale$ and $p_m$. A similar expression holds
for the Woods--Saxon potential well, the integration of the corresponding
exclusive cross--section extending over the same $(\cale, p_m)$ domain.

The limits on the missing momentum variable appearing in 
the above integrals are most easily deduced by setting $\cale^-=0$, 
which yields
\beq
 p^-_{m,min} = \frac{1}{1+M_\N/M_{\scriptscriptstyle{A-1}}}
\left(q-\sqrt{q^2 -\left(1+
\frac{M_\N}{M_{\scriptscriptstyle{A-1}}}\right) 
\left[q^2 - 2M_\N (\omega-E_S)\right]}\right)
\label{VII.5a} 
\eeq
and
\beq
 p^-_{m,max} =  \frac{1}{1+M_\N/M_{\scriptscriptstyle{A-1}}}
\left(q+\sqrt{q^2 -\left(1+
\frac{M_\N}{M_{\scriptscriptstyle{A-1}}}\right) 
\left[q^2 - 2M_\N (\omega-E_S)\right]}\right)
\label{VII.5b}
\eeq
and $\cale^+=0$, which yields:
\bea
 p^+_{m,min} &=& - p^-_{m,max}\qquad\quad{\mathrm {(always~ negative)}}
\label{VII.6a} \\
 p^+_{m,max} &=& - p^-_{m,min}\, .
\label{VII.6b}
\eea

In connection with these formulas we remind the reader that the 
scaling variable $y$ is customarily defined as the opposite of
the lower limit of the integration over the missing momentum variable. 
One thus sees that $y$ vanishes at $\omega={\widetilde\omega}$, being
positive (negative) for frequencies smaller (larger) than 
${\widetilde\omega}$.

We turn now to evaluate the $t$ inclusive cross--section 
in the uniform approximation for the distortion function $F$ [see
eqs. (\ref{V.14}) and (\ref{VI.16})]. In this case we get
\bea
&& \frac{d^2\sigma}{d\Omega_e d\epsilon'} =
\frac{1}{2\pi} \frac{M_\N}{ q}
\left\{ \theta\left(\omega-{\widetilde\omega}\right)
\left[\int_0^{p^+_{m,max}}p_m dp_m\int_{\cale^+}^{\cale^-} d\cale
+ \int_{p^+_{m,max}}^{p^-_{m,max}} p_m dp_m \int_0^{\cale^-}d\cale
\right]\right. +
\nonumber \\
&&\qquad\qquad \left.+ \theta\left({\widetilde\omega}-\omega\right)
\int_{p^-_{m,min}}^{p^-_{m,max}} p_m dp_m\int_0^{\cale^-} d\cale
\right\} 
\label{VII.7} \\
&&\qquad\times \frac{1}{|\qv-\pv_m|}
\int_0^{R_c} R^2 dR \left[\frac{d\sigma}{d\Omega_e}\right]_{sn}\!
\left.\left({\PR}{\hat p}-\qv, \qv\right)\right|_{\cos\theta=y_0} 
\theta\left(1-|y_0|\right)
\nonumber
\eea
where ${\PR}$ is again given by (\ref{VI.2}) and $y_0$ by (\ref{VI.16}),
while the upper limit of the 
integral over $R$ is found by solving the equation
\beq
 V(R_c) = \frac{(\qv-\pv_m)^2}{2M_\N} 
\label{VII.8}
\eeq
which, for the harmonic oscillator potential, yields
\beq
 R_c = \frac{1}{\omega_o}
\sqrt{\frac{1}{M_\N}}\left\{2\left[ V_o +(\omega -\cale - E_S)
\right] - \frac{ p_m^2}{M_{\scriptscriptstyle{A-1}}} \right\}\, .
\label{VII.9}
\eeq
Analogous expressions hold for the Woods--Saxon well.

\section{Results}
\setcounter{equation}{0}

In this section the predictions of our theory are numerically appraised.
We first consider the exclusive cross--sections. They are displayed
in Fig.~3a--d as a function of the missing momentum for various 
missing energies at $q=300$~MeV/c (Figs.~3a and 3b) and at $q=500$~MeV/c
(Figs.~3c and d); results both in the eikonal (Figs.~3a and c) and in 
the uniform
(Figs.~3b and d) approximation for the distortion of the outgoing
nucleon are shown. The energy transfer $\omega$ has been choosen
here to be close to the quasi--elastic peak ($\omega\simeq q^2/2M_\N+E_S$).
All the figures displayed in this first set refer to the
harmonic oscillator well, with $V_o=55$~MeV and 
$\hbar\omega_0=41/A^{1/3}\simeq 12$~MeV for the $^{40}$Ca nucleus.

The corresponding results for the Woods--Saxon well 
(in the same kinematical conditions and in 
the two approximations employed for the distortion operator) 
are shown in Figs.~4a--d; we use $V_1=49.8$~MeV, $R_o=1.2 A^{1/3}$~fm
and $a=0.65$~fm.

A few features are worth commenting:  one is related to the 
striking differences between the exclusive cross--sections evaluated
in the uniform and in the eikonal approximations. In the first case
the cross--sections are seen to be fairly constant over the whole
range of the kinematically allowed values of the missing momentum 
$p_m$, {\it independently} from the potential well binding the 
nucleons into the nucleus. In the second case the cross--sections show
a tendency to peak at low missing momenta and to fall off quite rapidly;
moreover they turn out to be restricted to a rather limited range
of $p_m$, much smaller than the one allowed by the pure kinematics. 
These outcomes clearly reflect the drastically different
distortion functions $F$ of the two cases.

A second feature refers to 
the exclusive cross--section in the eikonal approximation, which is quite 
sensitive to the shape of the potential well: indeed in this approximation
the outgoing particle keeps memory of the initial momentum and hence of
the spectral function inside the nucleus. The eikonal cross--sections 
turn out in fact to reflect quite closely the structure of $S(p,\cale)$, 
which is illustrated in the Appendix A.

The effects of the distortion operator can be even better appreciated by
comparing the above exclusive cross--sections with the corresponding
ones, evaluated in PWIA [see equation (\ref{pwia})]: this is done 
in Figs.~5a and 5b, for the 
harmonic oscillator and the Woods--Saxon well, respectively. In both 
cases $q=300$~MeV/c, $\omega$ is close to the quasi--elastic peak and 
$\cale=10$~MeV/c. The PWIA turns out to be closer, but still appreciably
different from the distorted 
cross--section evaluated in eikonal approximation, 
spanning a range of missing momenta ($p_m$) larger than in the eikonal
approximation itself, but much smaller than in the uniform approximation. 
The sensitivity to the shape of the mean field is strong also in PWIA.

Concerning the inclusive cross--sections, they are displayed 
in Figs.~6a and 6b at $q=200$~MeV/c and $q=500$~MeV/c, respectively, 
for both the harmonic oscillator and the Woods--Saxon well;
they are obtained via the integration in
the $(\cale, p_m)$ plane of the exclusive cross--section, using for the 
single nucleon cross--section $\sigma_{Mott}$ only. The calculation has 
been performed both  in
the uniform and in the eikonal approximations, which however cannot
be distinguished in the figures, owing to their identity. 
The expected 
numerical coincidence of the inclusive cross--sections obtained by
integrating the two, markedly different, exclusive cross--sections is thus
seen to be realized, providing that higher order 
relativistic effects in the single nucleon cross--section are ignored, 
which in turn amounts to keep only the Mott cross--section. 
When however the fully relativistic expression for the single nucleon 
cross--section, eq.(\ref{VII.10}), is employed, then the  above mentioned 
coincidence  between inclusive cross--sections no longer holds, especially 
near the maximum, but the discrepancy remains mild up to 
$q= 500$~MeV/c, as illustrated in Fig.~7a and 7b, which exhibit 
the same cases as in Fig.~6a and 6b. 

Also displayed in Figs.~6a and 6b are the
inclusive cross--sections obtained through
the polarization propagator method. These differ in the low energy 
side from the cross sections produced by the integration of the 
exclusive processes in the missing
energy--missing momentum plane: at low $q$ the difference is sizeable,
but it becomes negligible as $q$ increases. In general the inclusive 
cross sections obtained through the polarization propagator turn out
to be larger, as discussed in the 
introduction (we remind that the difference arises from the bound 
unoccupied orbits). The effect disappears at large $q$, as it should.

As a consequence of these findings, we confirm that little 
can be learned from the
inclusive electron scattering, treated within the mean field 
approximation, about the mechanism responsible for 
the distortion of the outgoing nucleon wave \cite{Lenz}.
Furthermore only a very weak dependence of
the inclusive cross--sections on the shape of the potential well is 
observed. 

Accordingly one is lead to conclude that while the 
nucleon--nucleon correlations ( of short and long range) in the initial state 
 affect both the inclusive and exclusive inelastic electron scattering, 
other details of the dynamics of the emitted nucleon, mainly reflecting
the shape of the mean field and how it affects the way of the nucleon out of the
nucleus, are more conveniently studied with exclusive processes.

\section{Conclusions}
\setcounter{equation}{0}

In this paper we have applied the semiclassical approach to the exclusive 
electron scattering, going beyond the PWIA. Indeed the
FSI has been accounted for in the mean field approximation including 
antisymmetrization. 

A few outcomes of the present study are worth to be recalled: the first one 
is the strong sensitivity of the exclusive process to the 
distortion of the outgoing nucleon wave: the stronger is the latter, the
weakest are the remnants of the nuclear mean field in the cross--section,
as it happens for the rather extreme situation described by the uniform
approximation. A ``weak'' distortion, like the one entailed by the 
eikonal approximation, yields cross--sections more strictly related to 
the spectral function and hence closer to the PWIA, as
it should: in this framework we have shown that the exclusive 
cross--sections are quite different in shape, whether we adopt the harmonic
oscillator or the Woods--Saxon well for the nuclear mean field.

A second interesting result relates to the strong correlation between 
the distortion mechanism and the domain, in the missin energy -- missing 
momentum plane, where the exclusive cross--section exists. We have
found that in the eikonal approximation this domain is quite restricted, 
whereas in the uniform one it essentially covers the whole 
kinematically allowed region.

Concerning the inclusive cross--section, obtained by integrating 
the exclusive ones in the appropriate domain of the $(\cale,p_m)$ plane,
we have shown that they 
are quite insensitive both to the distortion of the outgoing nucleon, 
as we prove also with an analytic calculation, and to the specific shape
of the mean binding field, in agreement with 
previous direct evaluations and measurements of the
quasi--elastic inclusive process. Yet we should mention a small 
difference between the inclusive cross--section obtained by integrating,
as discussed above, the exclusive ones, and the ones obtained by a direct
calculation based on the so--called polarization propagator. Indeed, in this
last instance, bound excited states are embodied, which are not allowed to
the outgoing nucleon in exclusive processes.
Finally we have observed, in the inclusive cross--section, a mild 
dependence upon the distortion mechanism when higher order relativistic
effects are taken into account in the expression of the single nucleon
cross--section: this discrepancy, which is quite small up to the largest
momenta ($500\div 600$~MeV/c) considered here, signals the
need of a fully consistent relativistic approach, both in the currents
and in the mean field utilized to describe the nucleon dynamics inside 
the nucleus.

We have found that the semiclassical method yields sensible results, 
but of course the validity of the scheme  can only be ultimately assessed
by testing it against both the experiment and fully quantum mechanical 
calculations.

Yet we consider the semiclassical 
method attractive by itself on three counts, namely:
\begin{enumerate}
\item for its simplicity,
\item for allowing to grasp the role of the FSI in a remarkably transparent 
way (this follows either by comparing the eikonal with the uniform 
approximations for the distortion or by a comparison with the PWIA),
\item for retaining in the exclusive process 
the simplicity of the Fermi gas model but taking into account the local
features of the nuclear mean field through the folding of Fermi gases at
different densities. Quantum mechanical interferences between different
points in the nucleus, not considered here, occur in the second and 
higher orders of the $\hbar$ expansion.

\end{enumerate}

With reference to point 2 we note that in our formalism the mechanism of
the distortion is embedded into the function $F({\hat p}_\N,{\hat p})$ 
of Section 5,
at variance with traditional (quantum mechanical) approaches, where it is
included in the distorted wave function of the outgoing nucleon. 

Clearly, before attempting to test the semiclassical approach against the 
experimental data, a realistic expression for the function $F$ should be 
developed: indeed the ones discussed in this paper were just meant as an 
illustration.

Concerning point 3 it is worth pointing out the correspondence with
the quantum mechanical approach of ref.\cite{REF7}, where a different
Fermi gas is associated with each individual orbit of the nucleons inside
the nucleus rather than with elementary volumes into which the nucleus 
is ideally split.

In the present work we have left out important dynamical effects 
arising from
\begin{enumerate}
\item the nucleon--nucleon correlations
\item the meson exchange currents
\item the two--step processes, which occur when the nucleon which has absorbed 
the photon scatters, in its way out, with another one in the nucleus 
or when a second 
nucleon is emitted from the excited daughter nucleus and comes out almost 
simultaneously to the one directly hit by the impinging photon. 
These mechanisms can of course contribute to both the $(e,e'p)$ and to 
the $(e,e'NN)$ exclusive cross--sections.
\end{enumerate}

Undoubtedly these processes can be, and in fact have partly been, treated 
in a fully quantum mechanical framework\cite{Benhar,Sick2}. 
Yet, as in the case of the FSI in the mean field
approximation, we believe that the semi--classical method can be 
advantageously employed in dealing also with this far from simple physics
to gain at least an orientation on its role in shaping the remarkably 
complex structure of the exclusive response in the missing energy--missing 
momentum plane.

\appendix

\section{\bf Appendix }

We calculate here explicitly the diagonal semi--classical spectral 
function for a few typical potential wells.

\begin{enumerate}
\item {\bf Square potential well}
\newline
Let's call $V_o (> 0)$ the depth of the well and $\bar R$ its range:
\beq
V(R)=\left\{
\begin{array}{ccc}
-V_o  \qquad\qquad &{\mathrm for}\quad &0\le R\le {\bar R}\\
0\qquad\qquad &{\mathrm for}\quad &R>{\bar R}
\end{array}
\right.
\eeq
Then one gets for the Fermi momentum, only defined for $R<\bar R$,
\beq
k_F = \sqrt{2 M^*_\N (\epsilon_F + V_o)}. 
\label{IV.14}
\eeq
An easy calculation yields then for the spectral function the following
expression
\beq
S_{sw} (p, {\cal E}) = \frac{4}{3} \pi \ \bar R^3 \ \theta \left( 
\sqrt{2 M^*_\N (\epsilon_F + V_o)} - p\right) \ \delta \left[ {\cale} -
\left(\epsilon_F + V_o - \frac{p^2}{2 M^*_\N}\right) \right]
\label{IV.15}
\eeq
which looks indeed like the FG one.

The support of (\ref{IV.15}) in the (${\cal E}, p$) plane is a line.
Note that
\beq
4 \int^{\infty}_0 \ d {\cal E} \ S_{sw} (p,{\cal E}) = 4 \ \frac{4}{3} \pi 
{\bar R}^3 \ \theta \left( \sqrt{2 M^*_\N (\epsilon_F + V_o)} - p\right)
\label{IV.16}
\eeq
provides the momentum distribution of the nucleons inside the nucleus, while
\beq
4 \int \frac{d \vec p}{(2\pi)^3} \ \int d{\cal E} \ S_{sw} (p,{\cal E}) =
\frac{4}{3} \pi \ {\bar R}^3 \ \frac{2k^3_F}{3 \pi^2} = A
\label{IV.17}
\eeq
[with $k_F$ given by (\ref{IV.14})] is the number of nucleons, the 
factor of $4$ accounting for the spin--isospin degeneracy. Of course 
the normalisation (\ref{IV.17})
holds valid if the Fermi energy fulfills the condition
\beq
\epsilon_F = 
\frac{1}{8} \ \frac{(9 \pi A)^{2/3}}{ M^*_\N {\bar R}^2} - V_o.
\label{IV.18}
\eeq

\item{\bf Harmonic oscillator potential well}

Setting
\beq
V(R) = \frac{1}{2} M^*_\N \omega_o^2 R^2 - V_o\, ,
\label{IV.19}
\eeq
$V_o$ being a positive constant of, say, $55$~MeV and $\omega_o$ the 
harmonic oscillator frequency, one easily derives the spectral function
\beq
S_{ho} (p,{\cal E}) = \sqrt{2} \ \frac{4 \pi}{(M^*_\N \omega^2_o)^{3/2}} \
\theta({\cale})\sqrt{\epsilon_F + V_o - p^2/2 M^*_\N - {\cale}} \,\,
\label{IV.20}
\eeq
whose support is no longer a line in the (${\cal E},\,p$) plane, but rather
a two--dimensional domain bound by the curve
\beq
{\cal E} = \epsilon_F + V_o - p^2/2 M^*_\N\, ,
\label{IV.21}
\eeq
the range of the allowed momenta being
\beq
0 \leq p \leq p_{ho}^{max}
\label{IV.22}
\eeq
with
\beq
p_{ho}^{max} \ = \ \sqrt{2 M^*_\N (\epsilon_F + V_o)}\, .
\label{IV.22'}
\eeq
Although (\ref{IV.21}) is reminiscent of the Fermi gas, the actual support of 
$S_{ho}$ is more closely related to the one of a quantum mechanical 
harmonic oscillator. Indeed in the latter case the spectral function is
substantially different from zero only in the region where $S_{ho}(p,
{\cal E})$ does not vanish, although it lives only on a set of parallel 
lines corresponding to the discrete harmonic oscillator eigenvalues.

In Fig.~8 the semi--classical harmonic oscillator spectral function is 
displayed for the case of $^{40}$Ca $(\omega_o \simeq 12$~MeV). 
At variance with the quantum mechanical situation where ${\cal E}$ is 
quantized and $p$ is a continuous variable, now both  ${\cal E}$ and $p$ are
continuous. Furthermore the semi--classical $S_{ho} \ (p,{\cal E})$ 
vanishes abruptly along the line (\ref{IV.21}) whereas in the quantum 
case there is a small leakage across this line. Finally the wild 
oscillations characterizing the quantum momentum distribution of the 
nucleons in a given shell are now replaced by a smooth behaviour with $p$.
By comparing this figure with Fig.8 of ref.\cite{REF7}, which is also 
obtained with an harmonic oscillator potential, one can better
appreciate the quantum mechanical versus semi--classical features of the 
spectral function. Yet
the semi--classical approach succeeds in filling up the region of the
$(\cale,p)$ plane where the strength of the mean field spectral function 
is expected to occur, 
keeping, to a large extent, the simplicity of the Fermi gas treatment, 
but without leading to the naive and extreme momentum distribution 
of the latter.

Indeed this quantity 
reads now 
\bea
&&4\int^\infty_0 d {\cal E} \ S_{ho} (p,{\cal E}) = 
\label{IV.23} \\
&&= 4 \sqrt{2} \ \frac{4 \pi}{(M^*_\N \omega^2_0)^{3/2}} \ 
\int^{{\cal E}_F + V_o - p^2/2 M^*_\N}_0 
d{\cal E} \sqrt{\epsilon_F + V_o - p^2/_{2 M^*_\N} - {\cal E}} 
\nonumber \\
&& =\frac{16 \pi}{3} \ \frac{1}{M^*_\N \omega^3_0} \ 
\left( 2 M^*_\N (\epsilon_F + V_o) - p^2\right)^{3/2} 
= \ \frac{16 \pi}{3} \ \frac{1}{M^{*3}_\N \omega^3_0} \ 
\left[\left(p_{ho}^{max}\right)^2 - p^2 \right]^{3/2}
\nonumber
\eea
with the normalization 
\bea
&&4 \int\frac{d \vec p}{(2 \pi)^3} \ d {\cal E}\, S_{ho} (p,{\cal E}) \ =
\ \frac{16 \pi}{3} \ \frac{1}{2 \pi^2} \ \frac{p_{h.o}^{max}}{(M^*_\N
\omega_0)^3} \ \int^1_0 \ dx \ x^2 (1 - x^2)^{3/2}
\nonumber \\
&&\quad = \frac{4}{3 \pi}\, \ \left( \frac{\left(p_{ho}^{max}\right)^2}
{M^*_\N \omega_0}\right)^3 \,
B\left(\frac{3}{2} \ , \ \frac{5}{2}\right) = 
\frac{2}{3} \ \left( \frac{\epsilon_F + V_o}{\omega_0}\right)^3
\label{IV.24}
\eea
where $B$ is the Euler function of second kind.\par

Since, according to the nuclear shell model, 
\beq
\omega_0 = \frac{41}{A^{1/3}} \ {\hbox{MeV}}\, ,
\label{IV.25}
\eeq
it follows that the momentum distribution is normalized to the
number of nucleons A providing the Fermi energy is fixed by
\beq
\epsilon_F = \left({3}{2}\right)^{1/3} 41 - V_o \simeq 47 - V_o.
\label{IV.26}
\eeq

For the sake of comparison, we illustrate in Fig.~9 the semi--classical
spectral function associated with the Woods--Saxon well, which reads:
\beq
S_{\scriptstyle WS}(p,\cale) =
4\pi R_{\delta}^2\theta\left[\epsF-\frac{p^2}{2M_\N}-V(R_{\delta})\right]
\frac{1}{|dV/dR|_{R=R_{\delta}}}
\label{IV.26bis}
\eeq
where
\beq
R_{\delta}=R_o+a\ln\left(\frac{V_1}{\cale+p^2/2M_\N-\epsF}-1\right)
\label{IV.26tris}
\eeq
and $dV/dR$ is the derivative of the Woods--Saxon well (\ref{VI.13a}).
The spectral function (\ref{IV.26bis}) has a different distribution of
strength with respect to the one obtained with the harmonic oscillator:
in particular its strength is quite sizeable close to the boundary 
of the region, in the $(\cale,p)$ plane, where it is confined;
on the other hand the latter is not much different from the one
of the harmonic oscillator well.
  
\item{\bf Power--law potential well}

For the potential
\beq
V(R) = - V_o + (\epsilon_F + V_o) \ \left(\frac{R}{R_c}\right)^n,
\label{IV.27}
\eeq
$V_o$ and $R_c$ being positive constants and $n = 1, 2, 3 ...$, one finds
\bea
S_{plw} (p,{\cal E}) &&= \frac{4}{3} \pi R^3_c \ 
\frac{3}{n (\epsilon_F + V_o)} \ \left( 1 - 
\frac{{\cal E} + p^2/2 M^*_\N}{\epsilon_F + V_o}\right)^{3/n-1} 
\nonumber \\
&&\qquad \times \theta \left[ \epsilon_F + V_o - \left({\cal E} +
\frac{p^2}{ 2 M^*_\N}\right)\right] 
\label{IV.28}
\eea
the Fermi energy being linked to the potential through
the relation (\ref{IV.2a}).

As in the previous instances $S_{plw} (p,{\cal E})$ lives in a domain
of the plane (${\cal E},p)$ bound by the curve
\beq
{\cal E} = \epsilon_F + V_o - \frac{p^2}{2 M^*_\N}
\label{IV.30}
\eeq
on which it vanishes, and within the range of momenta
\beq
0\leq p \leq p_{plw}^{max} 
\label{IV.31}
\eeq
with, as before,
\beq
p_{plw}^{max} = \sqrt{2 M^*_\N \ (\epsilon_F + V_o)}\, .
\label{IV.32}
\eeq
Note that for $n=\,3$ the spectral function is constant. Furthermore it becomes
closer and closer to the one of the square well (and, hence, of the FG) 
as $n$ becomes large.

The momentum distribution is given by
\beq
\int^\infty_0 \ d {\cal E} \ S_{plw} (p, {\cal E}) = 
\ \frac{16 \pi R^3_c}{3} 
\ \left(\frac{\epsilon_F + V_o - p^2 / 2M^*_\N}{\epsilon_F + V_o}
\right)^{3/n}
\label{IV.33}
\eeq
with normalization 
%
\bea
\int\frac{d\vec p}{(2\pi)^3} \int^\infty_0  d{\cal E} 
\, S_{plw}(p,{\cal E})
&=& \frac{8}{3 \pi} \ \left(R_c  p_{plw}^{max}\right)^3 
\, \int^1_0 \ dt \ t^2 (1 - t^2)^{3/n}
\nonumber \\
&=& \frac{4}{3} \ (p_{plw}^{max} R_c)^3 \ B\left( \frac{3}{2}, 
\frac{3}{n} + 1\right)\, ,
\label{IV.34}
\eea
and yields the nucleons' number $A$ when the Fermi energy is
fixed according to
\beq
\epsilon_F = \frac{1}{2 M^*_\N R^2_c} \ \left(\frac{3 \pi A}
{4B (3/2, 3/n + 1)}\right)^{2/3} - V_o\, .
\label{IV.35}
\eeq
Finally worth pointing out is that (\ref{IV.27}) interpolates between 
various forms of the mean field.

\end{enumerate}


\begin{figure}[p]
\caption[Fig.~\ref{fig1}]{\label{fig1}
The Feynman diagram representing the inclusive $(e,e')$ process.
}
\end{figure}         

\begin{figure}[p]
\caption[Fig.~\ref{fig2}]{\label{fig2}
Diagrammatic representation of an exclusive $(e,e'N)$ process.
}
\end{figure}         

\begin{figure}[p]
\caption[Fig.~\ref{fig3}]{\label{fig3}
a--d~~ Exclusive cross sections as a function of the missing momentum $p_m$
(in MeV/c) at $q=300$~MeV/c (a and b) and $q=500$~MeV/c (c and d); 
$\omega=q^2/2M_\N+E_S$. The eikonal (on the left) and the uniform (on the
right) approximations for the distortion operator are used; the harmonic 
oscillator potential is employed. In all figures curves 
corresponding to three different values
for the missing energy are displayed: $\cale=10$~MeV (continuous line),
$\cale=20$~MeV (dashed line) and $\cale=30$~MeV (dot--dashed line).
}
\end{figure}         

\begin{figure}[p]
\caption[Fig.~\ref{fig4}]{\label{fig4}
a--d~~ The same as in Fig.~3a--d. but employing the Woods--Saxon 
potential well.
}
\end{figure}         

\begin{figure}[p]
\caption[Fig.~\ref{fig5}]{\label{fig5}
a,b~~ Exclusive cross--sections as a function of the missing momentum 
$p_m$ (in MeV/c), for $q=300$~MeV/c, $\omega=q^2/2M_\N+E_S$, 
$\cale=10$~MeV, in the
eikonal (continuous line) and uniform (dashed line) DWIA, as well as in
PWIA (dot--dashed line). In (a) the harmonic oscillator potential is 
employed, in (b) the Woods--Saxon well.
}
\end{figure}         

\begin{figure}[p]
\caption[Fig.~\ref{fig6}]{\label{fig6}
a,b~~ Inclusive cross--sections at $q=200$~MeV/c (a) and 
$q=500$~MeV/c (b), as a function of the energy transfer $\omega$ (in MeV): 
the single nucleon cross section coincides with $\sigma_{Mott}$.
The continuous and dashed (coincident) lines refer to the integral of the 
exclusive cross--section in eikonal and uniform approximation, 
respectively, using the harmonic oscillator potential. The dot--dashed 
(eikonal) and dotted (uniform) lines are obtained with the Woods--Saxon 
well. The long--dashed line is the direct evaluation (with the harmonic 
oscillator potential) of the inclusive cross--section, through the 
polarization propagator.
}
\end{figure}         

\begin{figure}[p]
\caption[Fig.~\ref{fig7}]{\label{fig7}
a,b~~ The same as in Fig.~6a,b but using the fully relativistic single 
nucleon cross section (and omitting the direct calculation).
}
\end{figure}         

\begin{figure}[p]
\caption[Fig.~\ref{fig8}]{\label{fig8}
Semi--classical spectral function obtained with the harmonic oscillator
potential, as a function of missing energy and missing momentum.
}
\end{figure}         

\begin{figure}[p]
\caption[Fig.~\ref{fig9}]{\label{fig9}
Semi--classical spectral function obtained with the Woods--Saxon
potential, as a function of missing energy and missing momentum.
}
\end{figure}    

\begin{minipage}[p]{\textwidth}
\begin{center}
\mbox{\epsfig{file=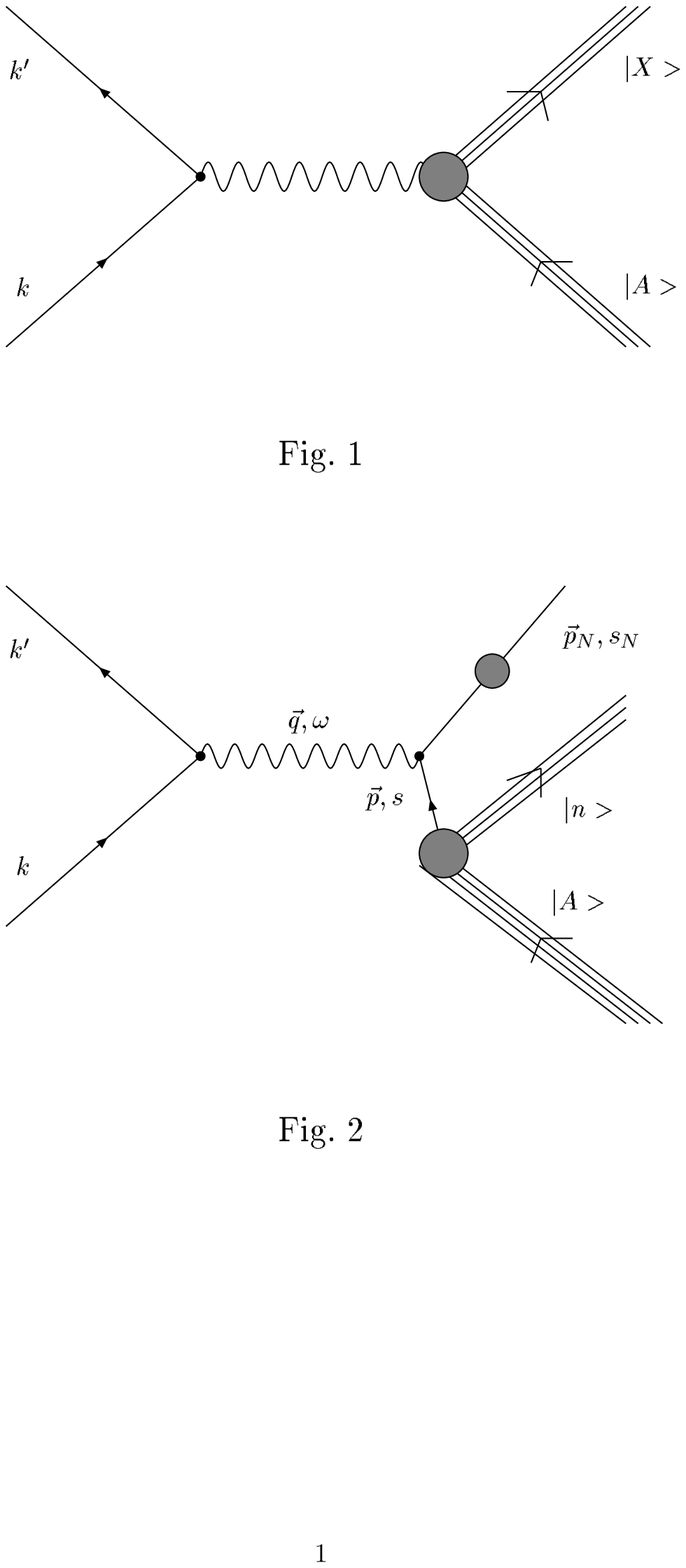,width=0.9\textwidth}}
\end{center}
\end{minipage}

\begin{minipage}[p]{\textwidth}
\begin{center}
\mbox{\epsfig{file=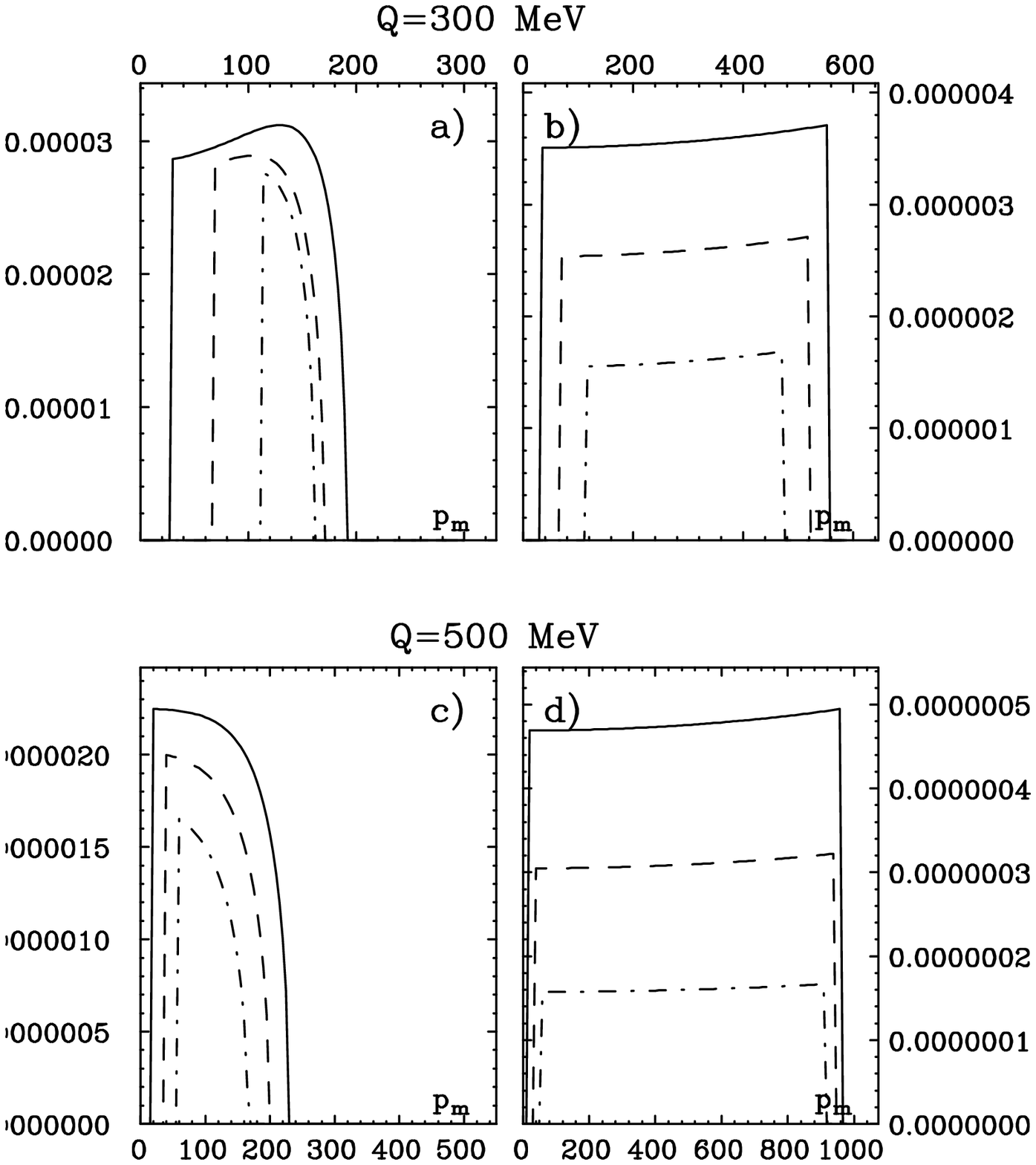,width=0.9\textwidth}}
\end{center}
\end{minipage}
\begin{center}
\Large Figure~\ref{fig3}
\end{center}     

\begin{minipage}[p]{\textwidth}
\begin{center}
\mbox{\epsfig{file=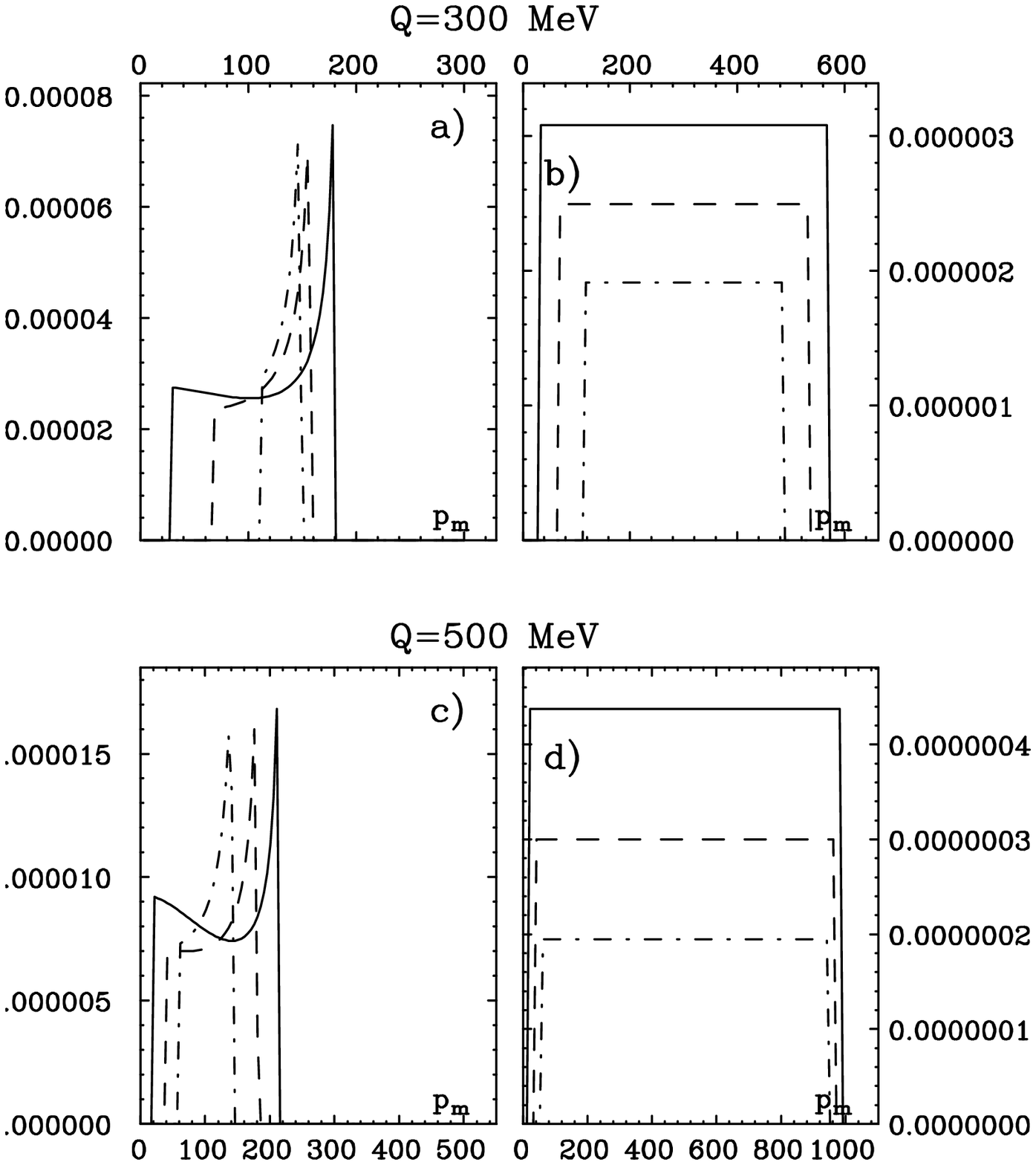,width=0.9\textwidth}}
\end{center}
\end{minipage}
\begin{center}
\Large Figure~\ref{fig4}
\end{center}     

\begin{minipage}[p]{\textwidth}
\begin{center}
\mbox{\epsfig{file=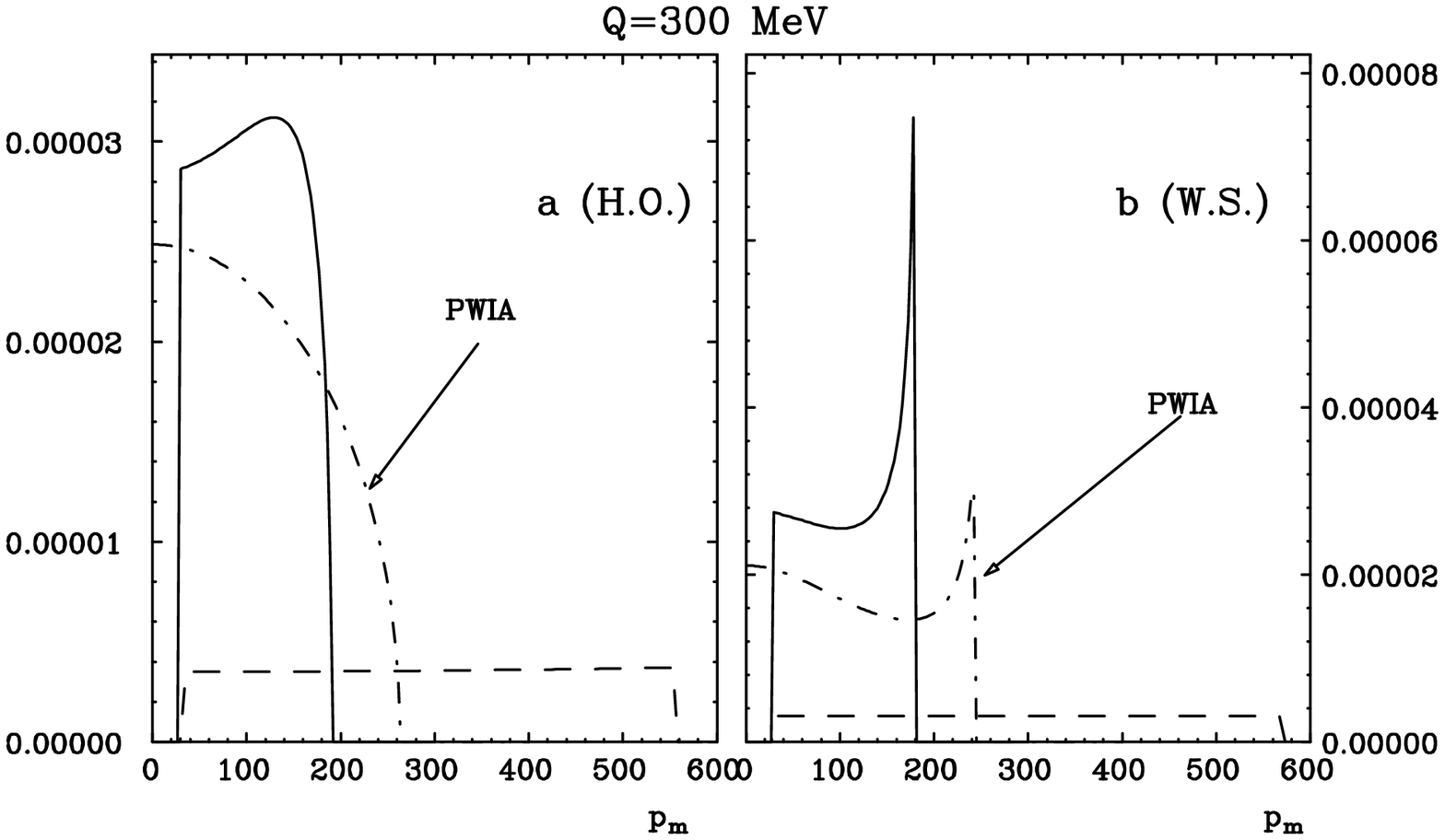,width=0.9\textwidth}}
\end{center}
\end{minipage}
\begin{center}
\Large Figure~\ref{fig5}
\end{center}     

\begin{minipage}[p]{\textwidth}
\begin{center}
\mbox{\epsfig{file=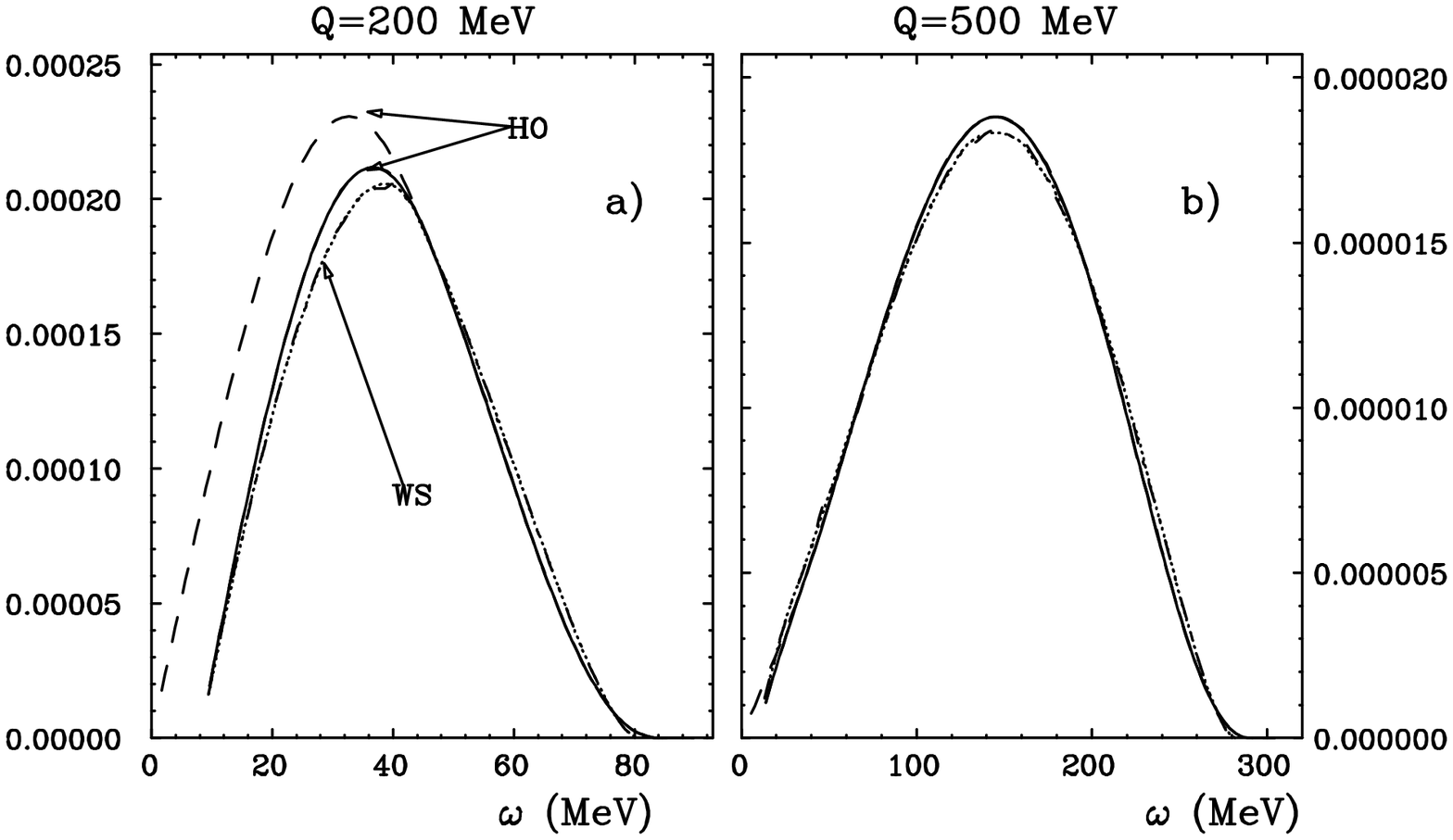,width=0.9\textwidth}}
\end{center}
\end{minipage}
\begin{center}
\Large Figure~\ref{fig6}
\end{center}     

\begin{minipage}[p]{\textwidth}
\begin{center}
\mbox{\epsfig{file=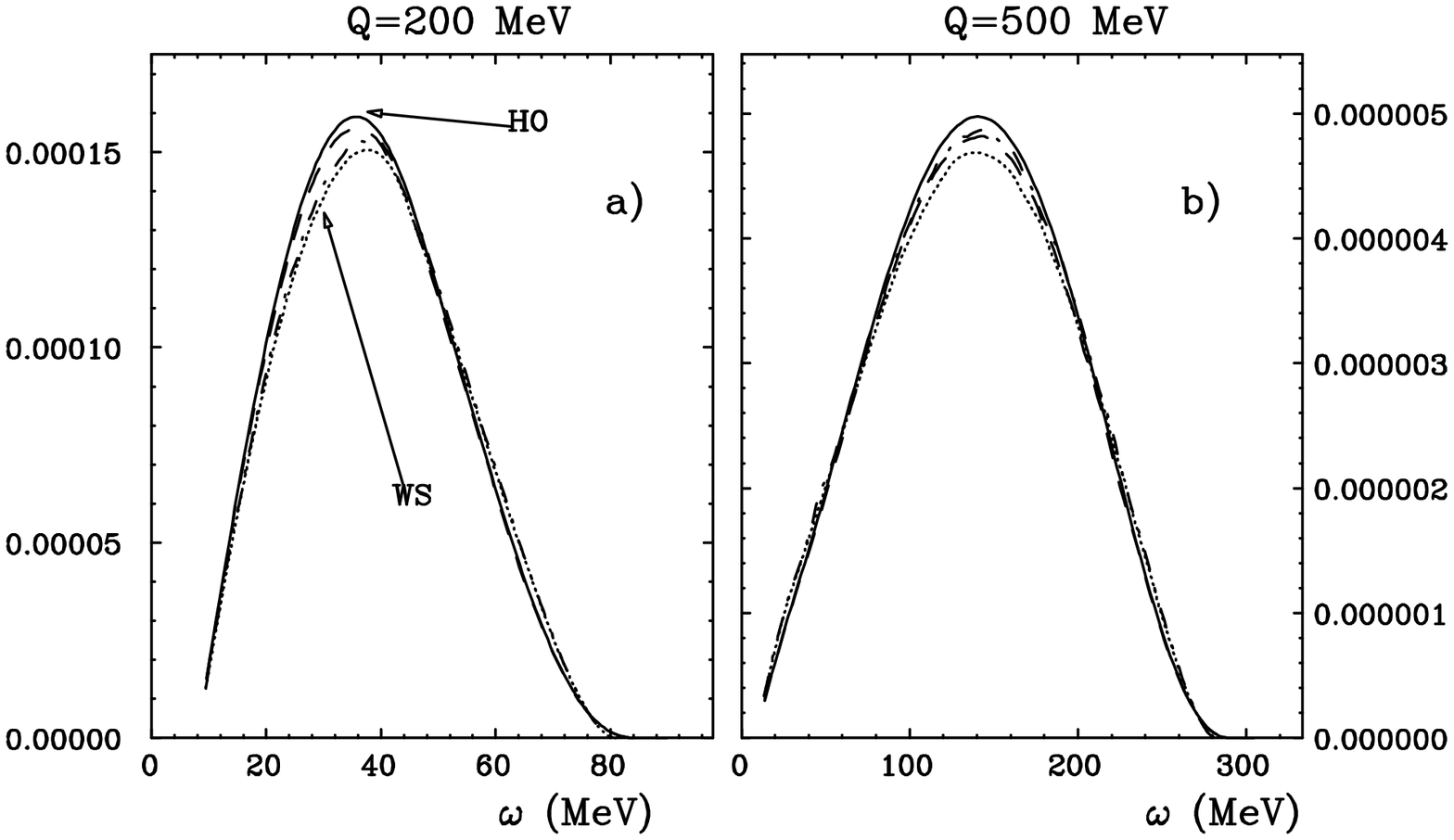,width=0.9\textwidth}}
\end{center}
\end{minipage}
\begin{center}
\Large Figure~\ref{fig7}
\end{center}     

\begin{minipage}[p]{\textwidth}
\begin{center}
\mbox{\epsfig{file=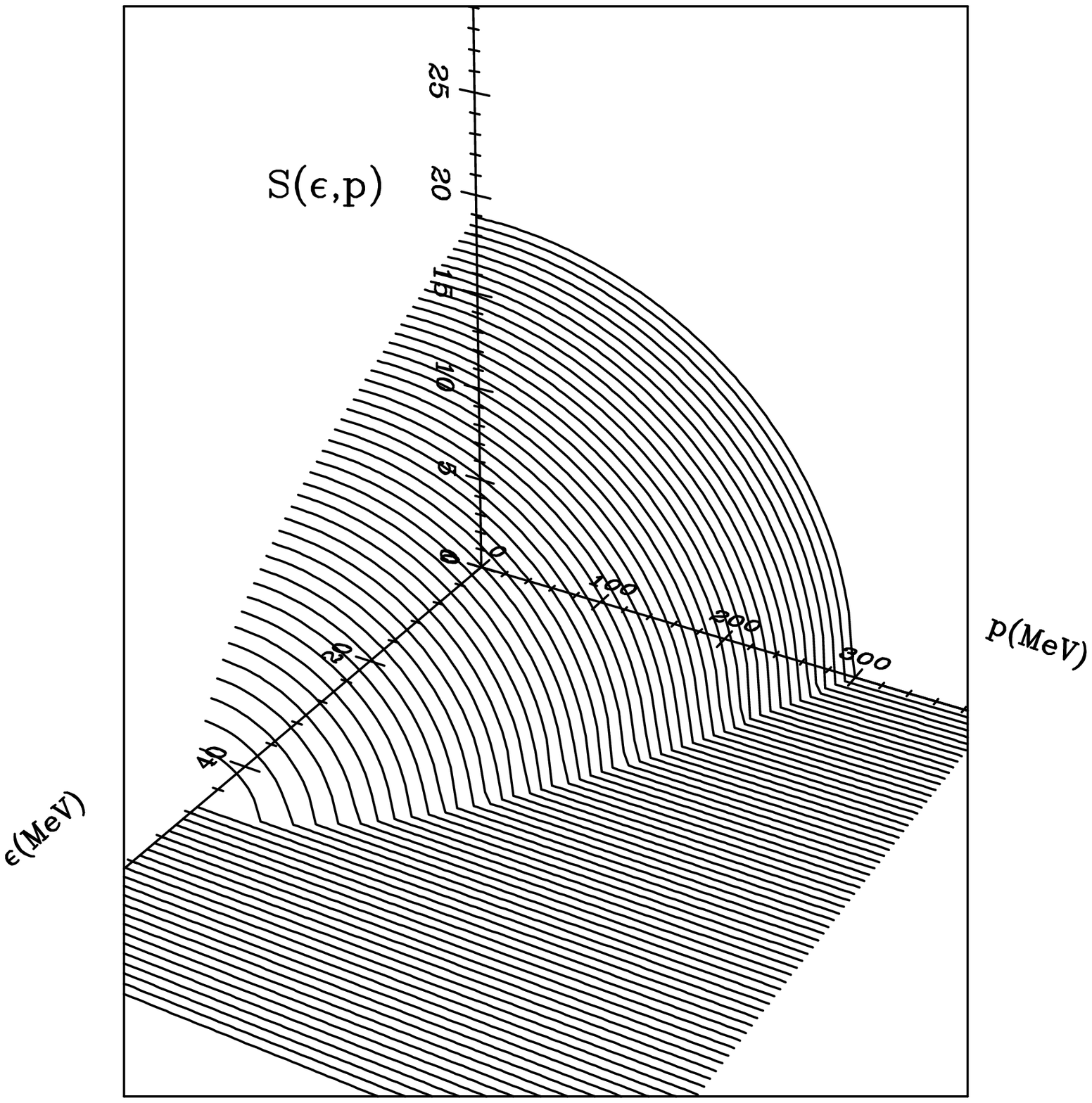,width=0.9\textwidth}}
\end{center}
\end{minipage}
\begin{center}
\Large Figure~\ref{fig8}
\end{center}     

\begin{minipage}[p]{\textwidth}
\begin{center}
\mbox{\epsfig{file=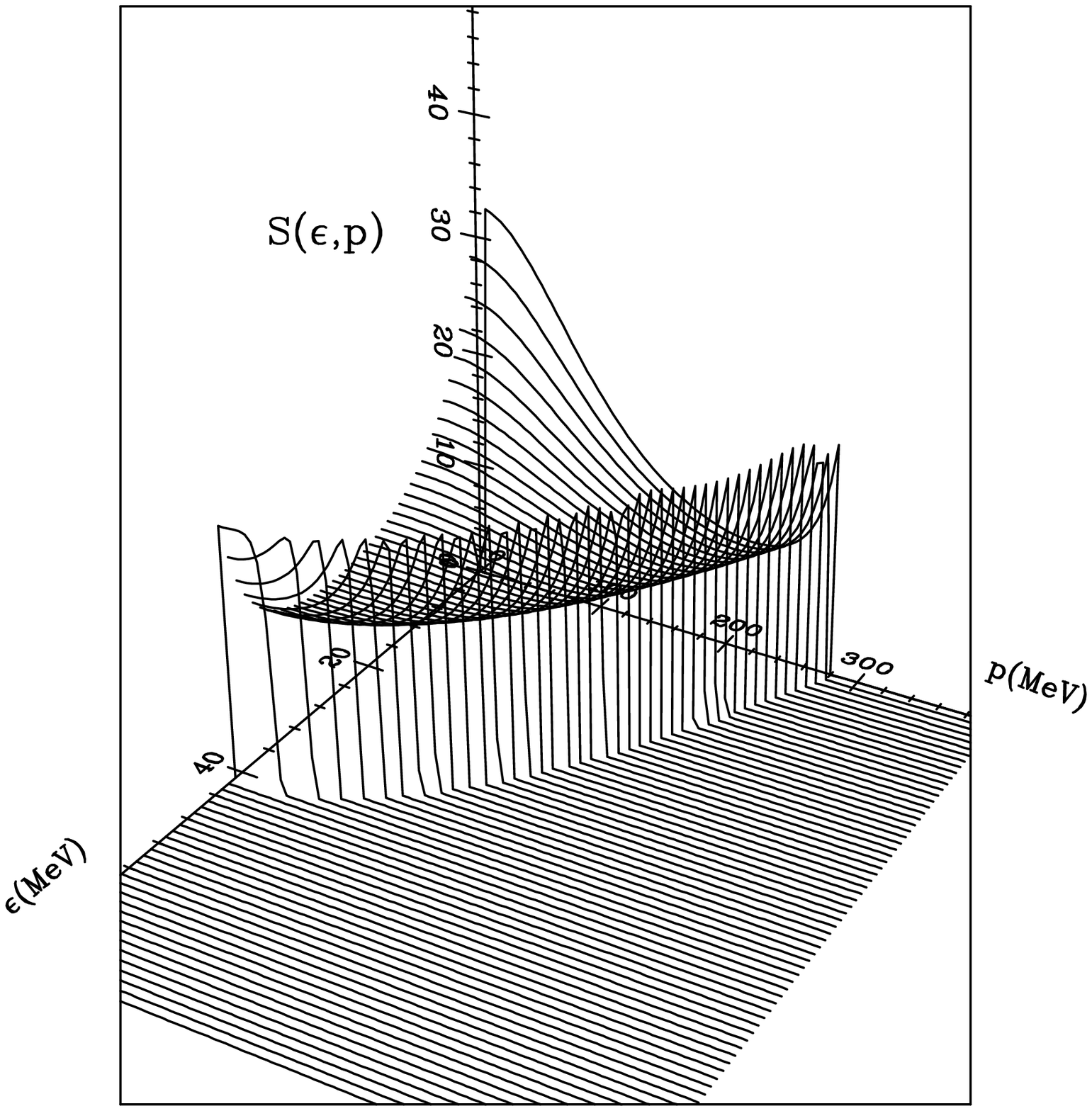,width=0.9\textwidth}}
\end{center}
\end{minipage}
\begin{center}
\Large Figure~\ref{fig9}
\end{center}     

\end{document}